\documentclass[
twocolumn,
aps,
prb,
reprint,
nonfootinbib,
superscriptaddress,
amsmath,amssymb,
]{revtex4-1}

\usepackage[pdftex]{graphicx, color}

\usepackage[pdftex,
            colorlinks=true,
            citecolor=blue,
            linkcolor=blue]{hyperref}
\usepackage{braket,ulem}
\usepackage{times,multirow,amsfonts,bm,xspace,pifont}
\usepackage{soul}
\usepackage{here}
\usepackage{physics}
\begin{document}
\newcommand{\rr}{{\bm r}}
\newcommand{\q}{{\bm q}}
\renewcommand{\k}{{\bm k}}
\newcommand*\wien    {\textsc{wien}2k\xspace}
\newcommand*\textred[1]{\textcolor{red}{#1}}
\newcommand*\textblue[1]{\textcolor{blue}{#1}}
\newcommand{\ki}[1]{{\color{red}\st{#1}}}
\newcommand{\sgn}{\mathrm{sgn}\,}
\newcommand{\GL}{{\mathrm{GL}}}
\newcommand{\talpha}{{\tilde{\alpha}}}
\newcommand{\tbeta}{{\tilde{\beta}}}
\newcommand{\mathN}{{\mathcal{N}}}
\newcommand{\mathQ}{{\mathcal{Q}}}
\newcommand{\bv}{{\bar{v}}}
\newcommand{\bj}{{\bar{j}}}
\newcommand{\zero}{{(0)}}
\newcommand{\one}{{(1)}}
\newcommand{\two}{{(2)}}
\newcommand{\three}{{(3)}}
\newcommand{\four}{{(4)}}

\newcommand{\YY}[1]{\textcolor{blue}{#1}}
\newcommand{\YYS}[1]{\textcolor{blue}{\sout{#1}}}

\newcommand{\add}[1]{\textcolor{red}{#1}}
\newcommand{\del}[1]{\textcolor{red}{\sout{#1}}}

\graphicspath{{./fig_submit/}}

\title{Supercurrent-induced antiferromagnetic order and spin-triplet pair generation \\ in quantum critical $d$-wave superconductors 
}

\author{Kyohei Nakamura}
\email[]{nakamura.kyohei.84w@st.kyoto-u.ac.jp}
\affiliation{Department of Physics, Graduate School of Science, Kyoto University, Kyoto 606-8502, Japan}

\author{Youichi Yanase}
\affiliation{Department of Physics, Graduate School of Science, Kyoto University, Kyoto 606-8502, Japan}

\date{\today}

\begin{abstract}
A supercurrent is well recognized as being of prime importance within mean-field theory, %and a quasiclassical theory 
but remains largely unexplored in strongly correlated electron systems (SCES) and the quantum critical region. 
To clarify the impact of the supercurrent on magnetism and superconductivity near an antiferromagnetic quantum critical point, we study the two-dimensional Hubbard model based on a fluctuation exchange approximation for a current-carrying superconducting state. %with a finite momentum of Cooper pairs.
We show a supercurrent-induced antiferromagnetism and emergence of spin-triplet Cooper pairs. The former results from Bogoliubov Fermi surfaces, suppression in the superconducting gap, and strong correlation effects beyond the mean-field theory.
Our results suggest that the supercurrent can bring out rich phenomena of superconductivity in SCES.
\end{abstract}

\maketitle

\section{Introduction}

A supercurrent is dissipationless and is of particular relevance to device applications such as superconducting electronics~\cite{Braginski2019-cp} and spintronics~\cite{Eschrig2015-jh,Linder2015-mc}.
The dissipationless current also enables us to analyze it within equilibrium states~\cite{Thinkam}, which are more tractable than nonequilibrium states with a normal electric current.
In addition, supercurrent-induced nonequilibrium transport can be investigated by the quasiclassical theory of superconductivity~\cite{SERENE1983221,Rammer1986}.
Then, theoretical studies on the supercurrent range from superconducting spintronics, such as spin transfer torque~\cite{Waintal2002,Zhao2008,Linder2011}, magnetization dynamics~\cite{Buzdin2008,Kulagina2014,Rabinovich2019}, non-collinear magnetic orders~\cite{Takashima2018}, and spin switching~\cite{Sun2024}, to nonreciprocal phenomena~\cite{Nagaosa2024}, such as magnetochiral anisotropy~\cite{Wakatsuki2017,Wakatsuki2018,Daido2024} and a superconducting diode effect~\cite{Yuan2022,Daido2022,Ili2022,He_2022}.
The nonreciprocal phenomena have been experimentally observed in various materials~\cite{Qin2017,Yasuda2019,Yuki2020,Ando2020,Narita2022,Bauriedl2022,Lin2022,Le2024,Mizuno2022,Qi2025}.
Moreover, the supercurrent has also been predicted to drive topological phase transitions~\cite{Takasan2022,Sumita2022}.
Therefore, the supercurrent provides a rich platform for exploring magnetism, transport phenomena, and topological phases.

Most of the theoretical studies mentioned above are based on the framework of a mean-field theory or a quasiclassical theory, eliminating quantum fluctuations and strong correlation effects.
Then, the effect of the supercurrent in strongly correlated electron systems (SCES) remains largely elusive, even though the superconducting diode effect has been experimentally observed in exemplary SCES, such as kagome superconductors~\cite{Le2024} and high-$T_{\rm c}$ cuprate superconductors~\cite{Mizuno2022,Qi2025} .
Solving this problem requires fundamental research on the supercurrent in SCES.

Understanding the effects of the supercurrent in SCES would also be meaningful in manipulating quantum phases.
Superconductivity in SCES is usually mediated by quantum fluctuations derived from other quantum phases, and those quantum phases and superconductivity can be controlled by tuning parameters.
Examples %of such superconductors 
include high-$T_{\rm c}$ cuprate superconductors, $\mathrm{Fe}$-based superconductors, %$\mathrm{Ce}$-, and $\mathrm{U}$-based 
and heavy fermion superconductors.
In high-$T_{\rm c}$ cuprate superconductors~\cite{Bednorz1986,Wu1987,Maeda_1988,Moriya01072000}, chemical doping converts the antiferromagnetic (AF) phase to the superconducting phase, which is considered to be mediated by AF fluctuations~\cite{Moriya01072000,YANASE20031}.
Similar phase diagrams can also be seen in Fe-based superconductors~\cite{Kamihara2008,HOSONO2015399} except for $\mathrm{FeSe}$~\cite{Fong2008,Shibauchi_FeSe_review}, while not only AF fluctuations but also orbital or nematic fluctuations are shown to be essential~\cite{Kuroki2008,Mazin2008,Kontani2010}.
In Ce-based heavy fermion superconductors~\cite{steglich1979,JACCARD1992475,Movshovich1996,WALKER1997303}, applied pressure changes the AF metal phase to the superconducting phase.
Moreover, $\mathrm{U}$-based heavy fermion superconductors~\cite{Saxena2000,Sheikin2001,Aoki2001,Levy2005,Huy2007,Aoki2009,Ran2019sts,Aoki2019,Ran2019mfs,Rosuel2023} and $\mathrm{CeRh_2As_2}$~\cite{Khim2021} exhibit magnetic-field-induced superconducting phases.
As exemplified above, the quantum phases in SCES are controlled by static parameters such as chemical doping, physical pressure, and magnetic field. Conversely, these tuning parameters highlight various aspects of superconductivity near a quantum critical point. 
Then, a tantalizing question is whether the supercurrent can become a novel tuning parameter for dynamic control.
The elucidation of this question would open up new avenues for research on superconductivity in SCES.

The aim of this paper is to elucidate the impact of supercurrent on magnetism and superconductivity near an AF quantum critical point.
To achieve this aim, we consider the two-dimensional Hubbard model as a fundamental model for SCES.
In Sec.~\ref{sec:model}, we formulate a Dyson-Gorkov equation~\cite{gor1958energy} and a fluctuation exchange (FLEX) approximation~\cite{Bickers1989,BICKERS1989206,Moriya01072000,YANASE20031} for this model with finite momentum Cooper pairs in order to study the supercurrent-carrying state with AF fluctuations and electron correlation effects.

The results of this paper show that the supercurrent can induce AF order and generate spin-triplet Cooper pairs, which are respectively discussed in Sec.~\ref{sec:magnetism} and Sec.~\ref{sec:superconductivity}. 
The supercurrent-induced AF order, the main result of this paper, is attributed to the appearance of Bogoliubov Fermi surfaces~\cite{Agterberg2017,Brydon2018}, suppression of the superconducting gap, and strong correlation effects.
This result suggests that the supercurrent enables us to %explore interesting aspects of 
control superconductivity and magnetism in SCES.
We give a brief summary of this paper in Sec.~\ref{sec:conclusion}.

\section{Formulation}
\label{sec:model}

In this section, we set up a model to investigate the supercurrent near an AF quantum critical point.
We start from the two-dimensional Hubbard model, which is a standard model of superconductivity in the vicinity of the AF quantum critical point:
\begin{align}
    H=\sum_{\bm{k}\sigma}\varepsilon(\bm{k})c^\dagger_{\bm{k}\sigma}c_{\bm{k}\sigma}+U\sum_{i}n_{i\uparrow}n_{i\downarrow}-h\sum_{i,\sigma}\sigma n_{i\sigma} ,  \label{eq:Hubbard model}
\end{align}
where $\bm{k}$, $\sigma=\uparrow,\downarrow$, and $i$ are index of momentum, spin, and site, respectively, $c^\dagger_{\bm{k}\sigma}$ ($c_{\bm{k}\sigma}$) is an electron creation (annihilation) operator with momentum $\bm{k}$ and spin $\sigma$, and $n_{i\sigma}=c^\dagger_{i\sigma}c_{i\sigma}$ is an electron density operator at site $i$ with spin $\sigma$.
In the first term of Eq.~\eqref{eq:Hubbard model}, we consider the square lattice and a tight-binding energy dispersion,  
\begin{align}
    \varepsilon(\bm{k})=-2t(\cos{k_x}+\cos{k_y})+4t'\cos{k_x}\cos{k_y}-\mu  , \label{eq:energy dispersion}
\end{align}
where $t$ and $t'$ represent nearest and next nearest neighbor hopping, respectively.
The chemical potential $\mu$ is chosen so that the number density of electrons is $n=0.85$.
In the second and third terms of Eq.~\eqref{eq:Hubbard model}, $U$ and $h=\frac{1}{2}g\mu_BH$ are an onsite Coulomb repulsion and the Zeeman field, respectively.
Here, an orbital effect from magnetic fields is ignored for simplicity.
We set the nearest neighbor hopping $t=1$ as a unit of energy.

Normal and anomalous Green functions, $G(k)$ and $F(k)$, satisfy the Dyson-Gorkov equation~\cite{gor1958energy} as follows,
\begin{align}
     \tilde{G}(k; \bm{p})&=
     \begin{pmatrix}G_{\uparrow}(k; \bm{p}) & F(k; \bm{p})\\
     F^\dagger(k; \bm{p}) & -G_{\downarrow}(-k; \bm{p})  \end{pmatrix} \notag
     \\
     &=\begin{pmatrix}G^n_{\uparrow}(k;\bm{p})^{-1} & \Delta(k; \bm{p})\\
     \Delta^\dagger(k; \bm{p}) & -G^n_{\downarrow}(-k;\bm{p})^{-1} \end{pmatrix}^{-1} , \label{Dyson-Gorkov eq}
\end{align}
where
\begin{align}
    G^n_{\sigma}(k;\bm{p})^{-1}=i\omega_n-\varepsilon(\bm{k}+\bm{p})+\sigma h-\Sigma_{\sigma}(k;\bm{p}) ,
\end{align}
has the form of the Green function in the normal state. 
We introduced a shorthand notation as $k=(\bm{k},i\omega_n)$, and $\Sigma_{\sigma}(k;\bm{p})$ and $\Delta(k;\bm{p})$ are normal and anomalous self-energies, respectively. In contrast to the previous studies on static superconducting states with vanishing supercurrent~\cite{Moriya01072000,YANASE20031}, Cooper pairs with a momentum $2\bm{p}$ are assumed. 
We regard the momentum of Cooper as a parameter %pairs is used instead of 
corresponding to the supercurrent.
This is justified as long as the free energy as a function of $\bm{p}$ is downward convex.
We choose $\bm{p}$ to be compatible with the periodic boundary condition,
\begin{align}
    \bm{p}=\frac{2\pi}{L}(n_x,n_y) , \label{eq:momentum of Cooper pairs}
\end{align}
and assume that the supercurrent is applied along the $[100]$- or $[110]$-axis.
In the following formulation, functions of $k$ parameterized by $\bm{p}$ such as $A(k;\bm{p})$ are abbreviated as $A(k)$ for simplicity unless otherwise stated.

To determine the normal and anomalous self-energies, some approximations are necessary.
In this study, we adopt the FLEX approximation, %is adopted, including AF fluctuations. The FLEX approximation 
which is a conserving approximation based on the Luttinger-Ward formalism~\cite{Luttinger1960}.
The normal and anomalous self-energies are determined by the Luttinger-Ward generating function $\Phi[G_\sigma,F,F^\dagger]$ as follows,
\begin{align}
    \Sigma_{\sigma}(k)&=\fdv{\Phi}{G_{\sigma}(k)}, \\
    \Delta(k)&=-\fdv{\Phi}{F^\dagger(k)}, \\
    \Delta^\dagger(k)&=-\fdv{\Phi}{F(k)}.
\end{align}
In the FLEX approximation, the generating function consists of ladder and bubble skeleton diagrams and is given by,
\begin{align}
    \Phi[G_{\sigma},F,F^{\dagger}]=&\sum_{q}\left\{\mathrm{log}\qty[1-U\chi_{\pm}^0(q)] \right. \notag \\
    &\left. + \frac{1}{2}\mathrm{log}\qty[\qty{1-U\chi_F^0(q)}^2-U^2\chi_{\uparrow}^0(q)\chi_{\downarrow}^0(q)] \right. \notag \\ 
    &\left. + \frac{1}{2}U^2\qty[\chi_F^0(q)^2+\chi_{\uparrow}^0(q)\chi_{\downarrow}^0(q)] \right. \notag \\
    &\left. + U\qty[\chi_{\pm}^0(q)+\chi_F^0(q)]\right\} ,
\end{align}
where irreducible susceptibilities are given by
\begin{align}
    \chi^0_{\pm}(q)&=-\sum_{k}\qty[G_\uparrow(k+q)G_\downarrow(k)+F(k+q)F^{\dagger}(-k)] , \\
    \chi_{\sigma}^0(q)&=-\sum_{k}G_{\sigma}(k+q)G_{\sigma}(k) , \\
    \chi_F^0(q)&=-\sum_{k}F^{\dagger}(k+q)F(k) .
\end{align}
Here, $\sum_k$ and $\sum_q$ mean $T/L^2\sum_{\bm{k},\omega_n}$ and $T/L^2\sum_{\bm{q},\Omega_n}$, respectively, with $\omega_n$ $(\Omega_n)$ being the fermionic (bosonic) Matsubara frequency.
%$T$ and $L^2$ are the temperature and the system size.
We set the temperature $T=0.005<T_{\rm c} \simeq 0.015$ to be lower than the superconducting critical temperature $T_{\rm c}$.
The system size is set to $L^2=256\times256$.
For Matsubara frequencies, sparse-ir is adopted for efficient numerical calculations~\cite{Shinaoka2017,LiJia2020,Shinaoka2022}.

\section{results and discussion}
\label{sec:results and discussion}

\begin{figure*}[htbp]
    \centering
    \begin{tabular}{ccccc}
    (a)  $(n_x,n_y)=(0,0)$ &(b) $(n_x,n_y)=(1,1)$&(c) $(n_x,n_y)=(2,2)$&(d) $(n_x,n_y)=(1,0)$&(e) $(n_x,n_y)=(2,0)$\\ 
     \includegraphics[width=0.19\textwidth,height=0.114\textheight]{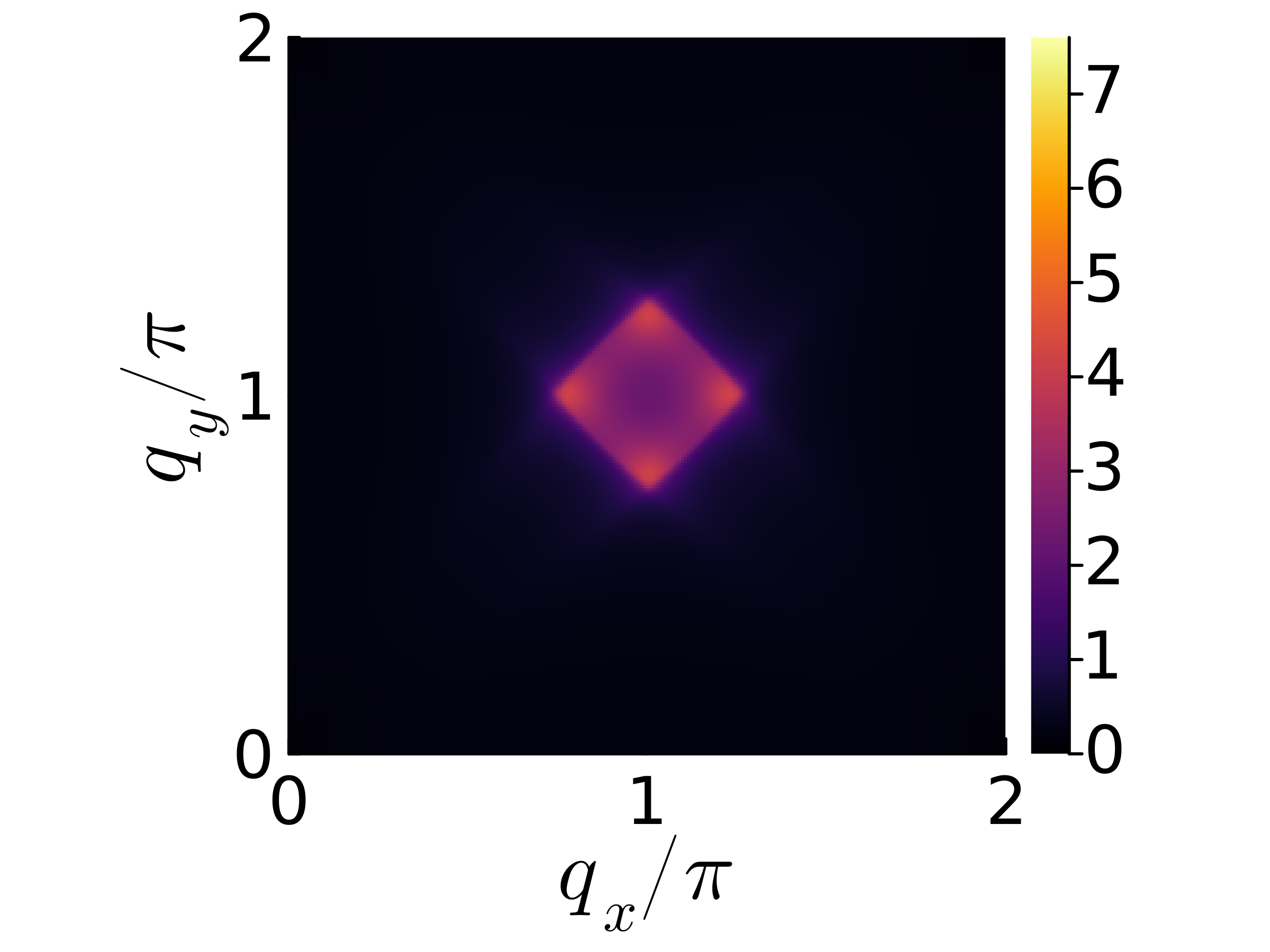} 
    &\includegraphics[width=0.19\textwidth,height=0.114\textheight]{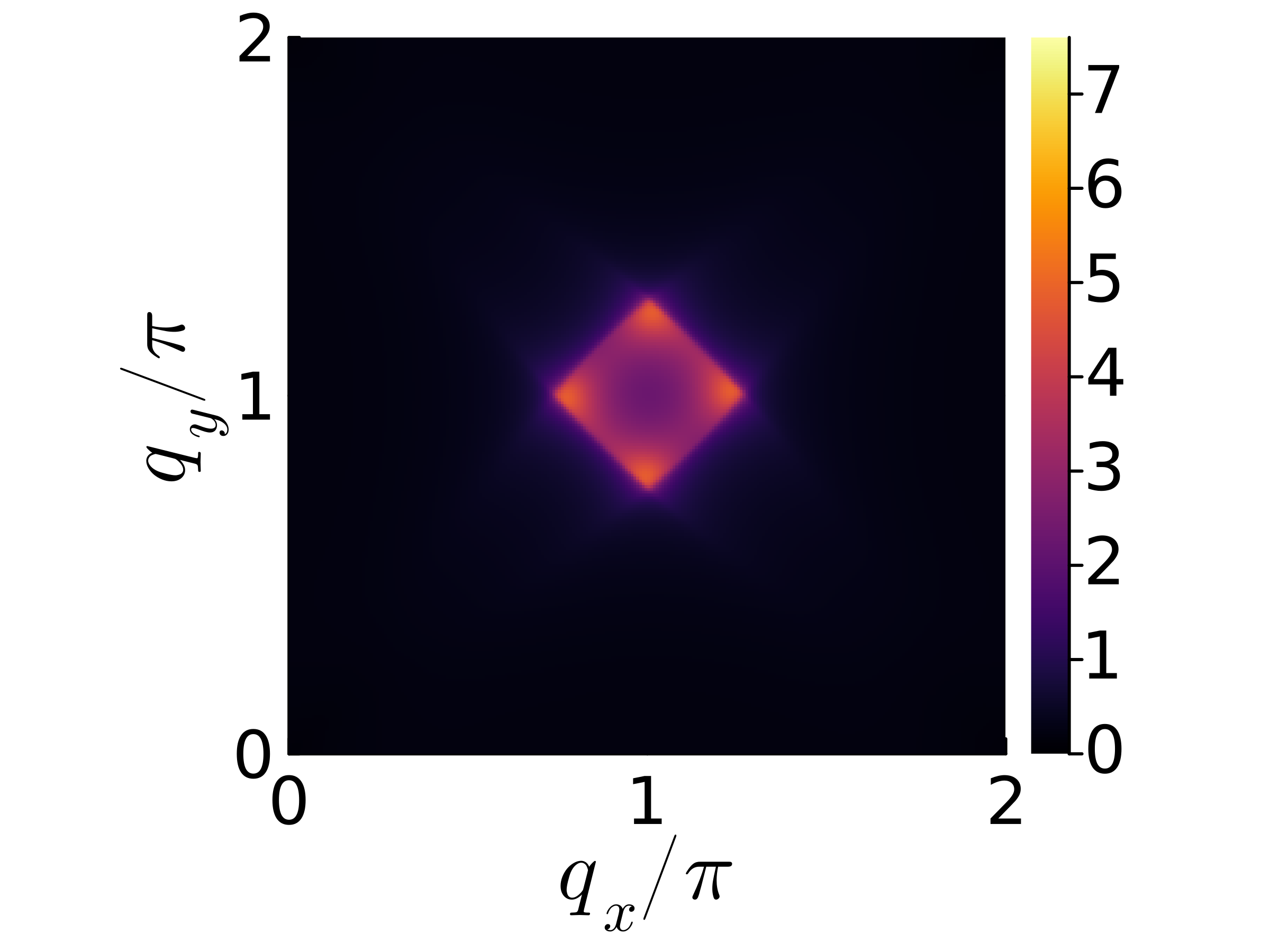}
    &\includegraphics[width=0.19\textwidth,height=0.114\textheight]{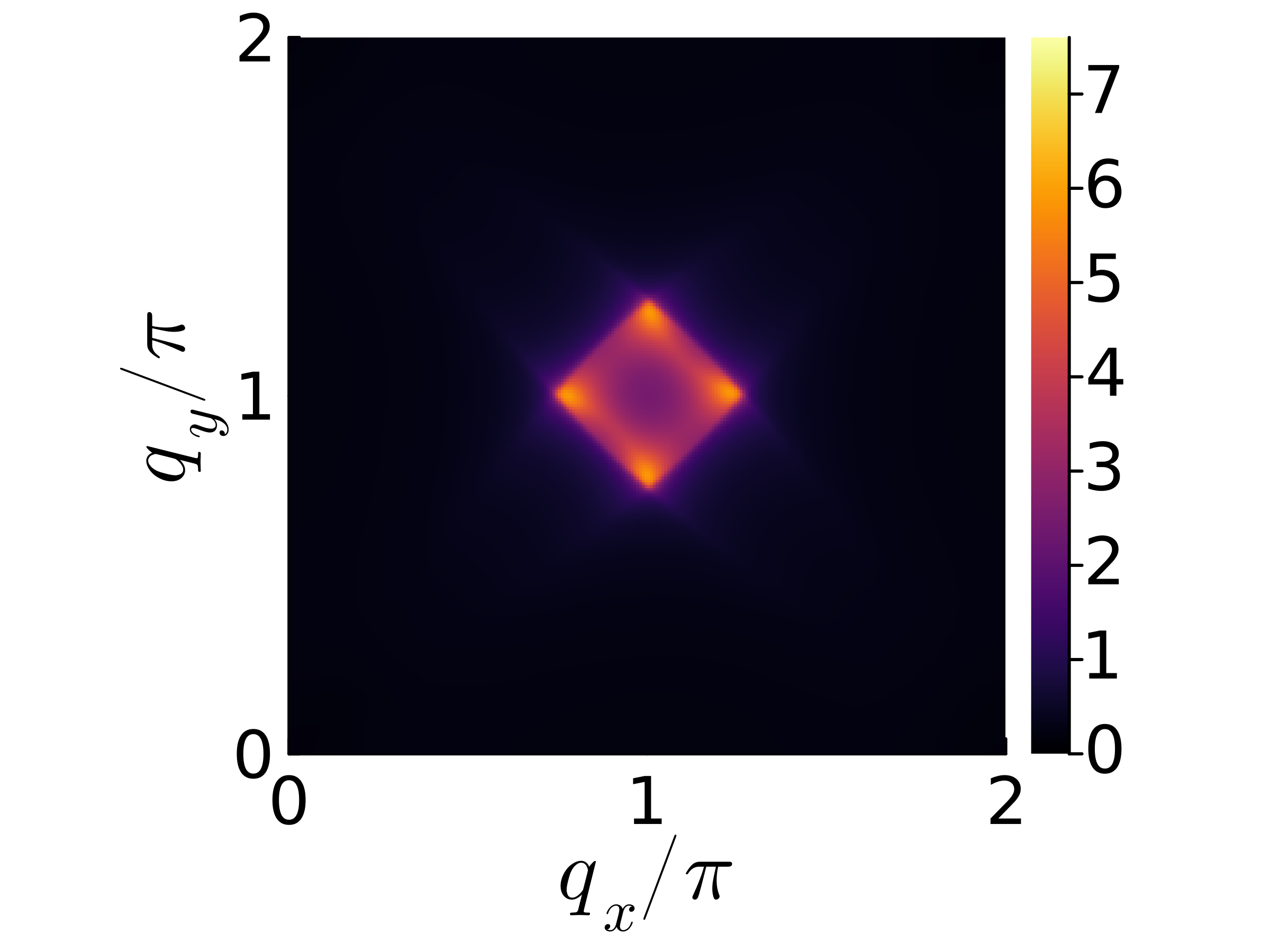}
    &\includegraphics[width=0.19\textwidth,height=0.114\textheight]{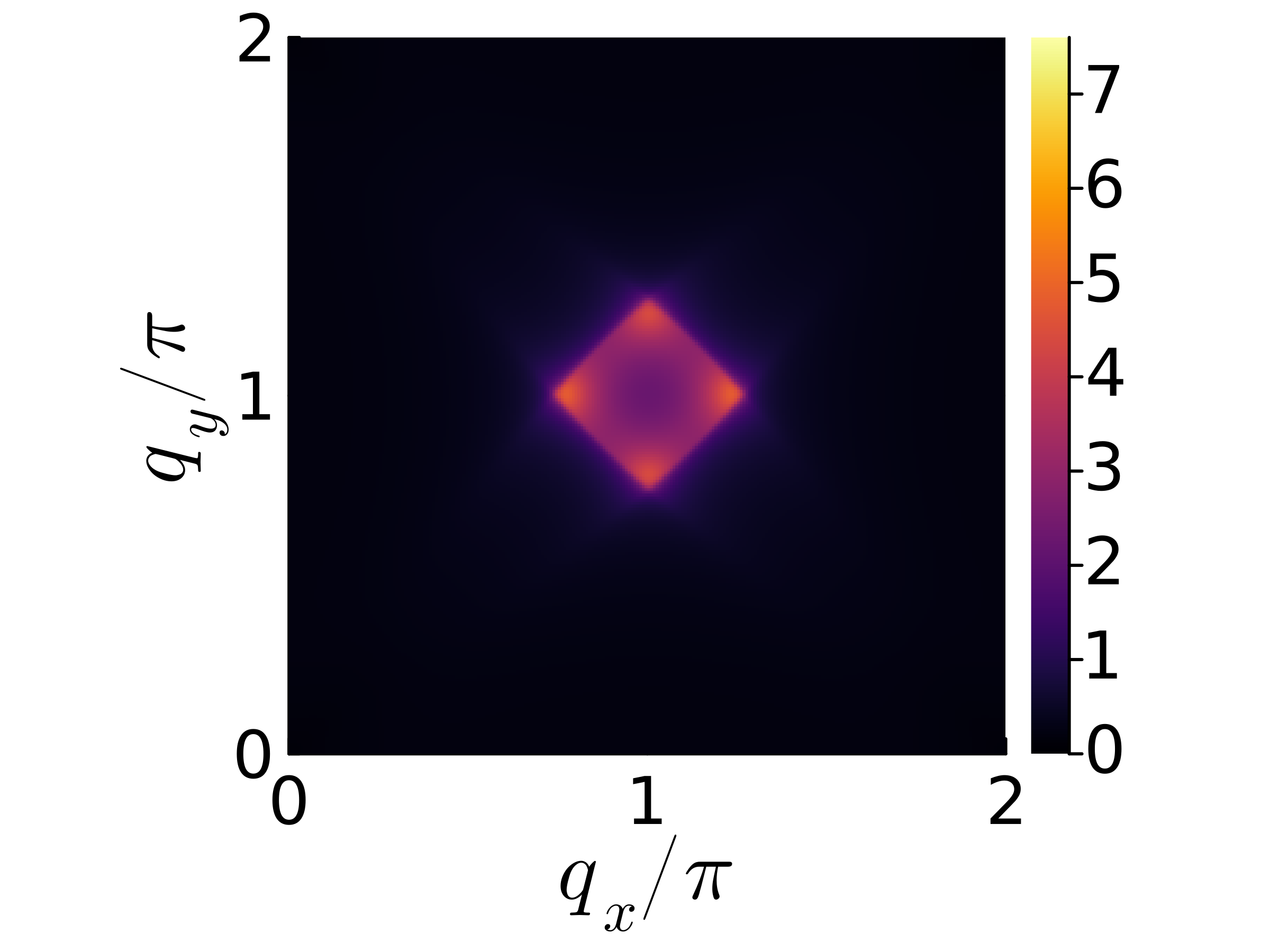}
    &\includegraphics[width=0.19\textwidth,height=0.114\textheight]{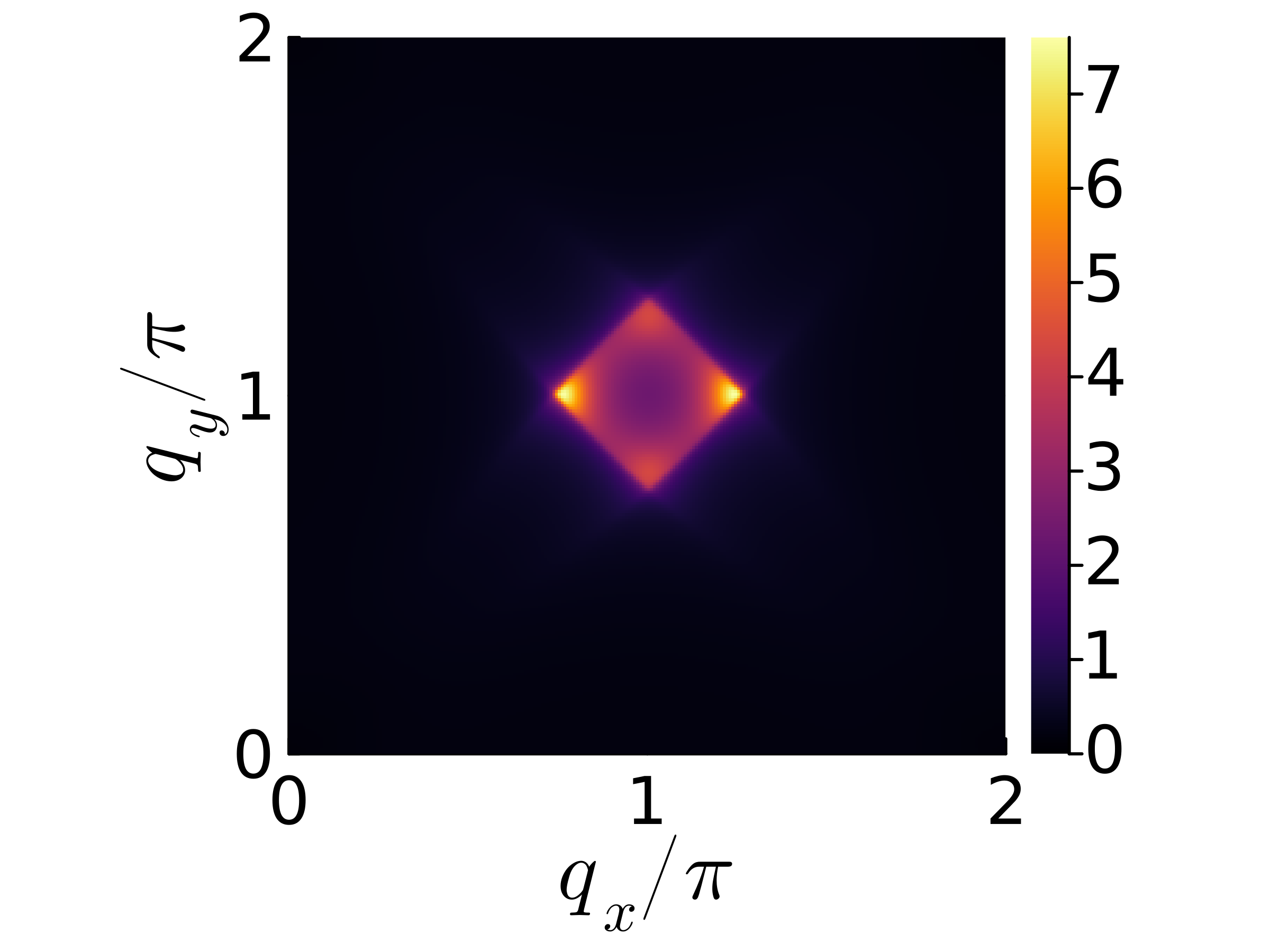} 
    \end{tabular}
    \caption{Momentum dependence of the static spin susceptibility, $\chi^\perp(\bm{q},i\Omega_n=0)$, in the (a) absence or (b)-(e) presence of the supercurrent. 
    The supercurrent is represented by the momentum of Cooper pairs, which is proportional to $(n_x,n_y)$.
    The parameters are set to $(t',U)=(0.25,3.5)$.
     }
    \label{fig:spin sus heatmap}
\end{figure*}

In the following results and discussion, the momentum of Cooper pairs is represented by integers $(n_x,n_y)/\!/\bm{p}$ [see Eq.~\eqref{eq:momentum of Cooper pairs}]. %supercurrent are represented by the finite meqomentum of Cooper pairs,
As mentioned in Sec.~\ref{sec:model}, we can represent the supercurrent by these parameters as long as the free energy as a function of $\bm{p}$ is downward convex.
In this region, the supercurrent is proportional to the Cooper pair's momentum ${\bm J} = 2 D_{\rm s} \,{\bm p}$ for small ${\bm p}$, with the superfluid density $D_{\rm s}$.
In contrast, the Cooper pair's momentum is not simply related to the supercurrent in the Fulde-Ferrell-Larkin-Ovchinnikov (FFLO) state~\cite{Fulde1964,Larkin1964}, because the stable superconducting state acquires a finite momentum ${\bm p}$ in the absence of the supercurrent. 
In the two-dimensional Hubbard model, the FFLO state stabilizes in the high magnetic field region, $h\gtrsim T_{\rm c}$~\cite{Vorontsov2005,Yanase2008}.
Although the relationship between the FFLO state and magnetism~\cite{Lortz2007,Young2007,Kenzelmann2008,Yanase2009,Sumita2023,Amin2024} is a fascinating subject, the main topic of this study is not the FFLO state.
Therefore, in this study, we address the zero and low field region, $h\lesssim T_{\rm c}\sim0.015$, to avoid the complexity of the FFLO state.

The magnetic field breaks isotropy in the spin space and renders transverse and longitudinal spin susceptibilities inequivalent.
In the two-dimensional Hubbard model above $T_{\rm c}$, the longitudinal AF spin fluctuation is suppressed by field-induced uniform magnetization, whereas the transverse AF spin fluctuation is enhanced~\cite{Sakurazawa2005}.
Although we have confirmed this field-induced anisotropy in the superconducting state, these changes are very small in the low magnetic field region. 
Therefore, the magnetic field is set to zero, $h=0$, in Sec.~\ref{sec:magnetism}. 
However, the cooperation of the magnetic field and the supercurrent impacts superconductivity and induces spin-triplet Cooper pairs. 
%Then, the low field has little effect on magnetism but superconductivity.
Thus, we take into account small but finite magnetic fields in Sec.~\ref{sec:superconductivity}.

\subsection{Supercurrent-induced magnetism}
\label{sec:magnetism}

In this subsection, we discuss an AF order by calculating the spin susceptibility in the presence of the supercurrent. This part is the main result of this paper. 
Transverse and longitudinal spin susceptibilities, $\chi^\perp(q)$ and $\chi^\parallel(q)$, are given by
\begin{align}
    \chi^\perp(q)&=\frac{\chi_{\pm}(q)+\chi_\mp(q)}{4}, \\
    \chi^\parallel(q)&=\frac{\chi_\uparrow(q)+\chi_\downarrow(q)}{4}+\frac{U\chi_\uparrow(q)\chi_\downarrow^0(q)}{2} \notag \\
                        &+\frac{\chi_F(q)\qty(1-U\chi_F^0(q))}{2},
\end{align}
where
\begin{align}
    \chi_{\pm}(q)&=\frac{\chi^0_{\pm}(q)}{1-U\chi^0_{\pm}(q)} , \\ 
    \chi_\mp(q)&=\chi_\pm(-q) , \\
    \chi_\sigma(q)&=\frac{\chi_\sigma^0(q)}{\qty(1-U\chi_F^0(q))^2-U^2\chi_\uparrow^0(q)\chi_\downarrow^0(q)}, \\
    \chi_F(q)&=\frac{\chi_F^0(q)}{\qty(1-U\chi_F^0(q))^2-U^2\chi_\uparrow^0(q)\chi_\downarrow^0(q)}.
\end{align}
The Stoner factor is defined as
\begin{align}
    \alpha_{S}(q)=U\chi^0_{\pm}(q) .
\end{align}
In this paper, a criterion, $\max\alpha_S(\bm{q},i\Omega_n=0) \geq 0.98$, determines the AF order, as in previous studies of two-dimensional models~\cite{YANASE20031}.
As mentioned at the beginning of this section, the calculations in this subsection are carried out at the zero magnetic field $h=0$, resulting in equivalence between the transverse and longitudinal spin susceptibilities.
Then, we address only the transverse spin susceptibility. %in this subsection.
In this case, only the spin-singlet order parameter of superconductivity appears even with the supercurrent.
Spin-triplet pair generation in finite magnetic fields is discussed in Sec.~\ref{sec:superconductivity}.

%Before discussing our results, we comment on an important parameter in spin susceptibilities.
%The second-neighbor hopping integral, $t'$, controls Fermi surface nesting, which plays an essential role in AF fluctuations.
%For example, $t'=0$ or perfect Fermi surface nesting means that the AF fluctuation diverges, although the peak of the AF fluctuation is too local to stabilize superconductivity.
%Then, moderate Fermi surface nesting is needed for superconductivity, and thus we choose the parameters $(t',U)=(0.25,3.5),(0.35,5.5)$.
%The Coulomb repulsion on site $U$ is chosen so that the superconductivity is stable enough to apply the supercurrent.

\begin{figure}[htbp]
    \centering
    \begin{tabular}{l}
    \includegraphics[width=0.4\textwidth,height=0.2\textheight]{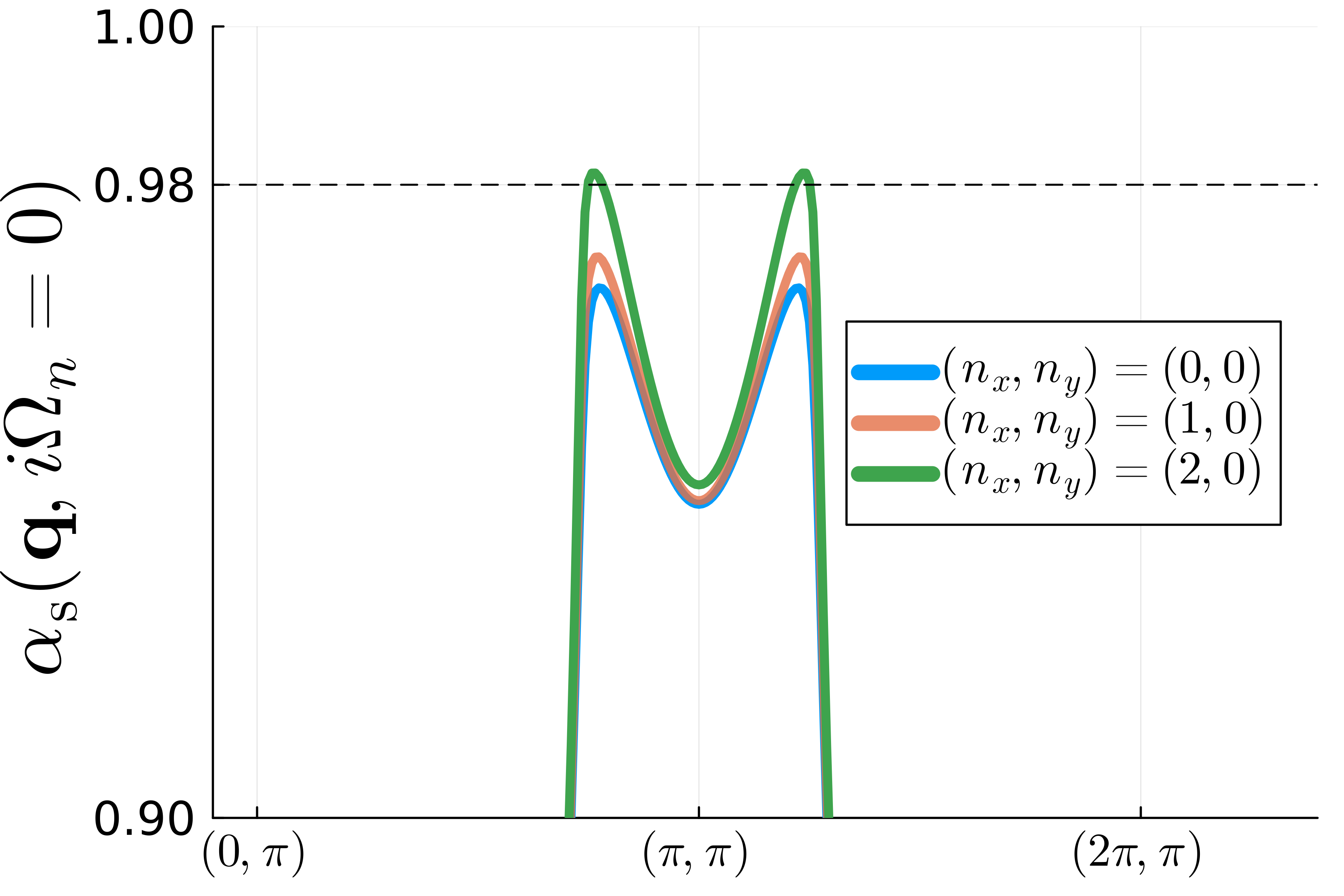}  
    \end{tabular}
    \caption{ Stoner factor, $\alpha_S(\bm{q},i\Omega_n=0)$, for momentum along the high symmetry line. 
    We choose the parameters $(t',U)=(0.25,3.5)$, the same as in Fig.~\ref{fig:spin sus heatmap}.
    The blue, orange, and green lines respectively show the Stoner factor for $(n_x,n_y)=(0,0),(1,0),$ and $(2,0)$.
    %The momentum of Cooper pairs, $(n_x,n_y)$, is used instead of the supercurrent.
    The Stoner factors for $(n_x,n_y)=(1,1)$ and $(2,2)$ have peaks slightly shifted from high symmetry lines and thus are not shown to avoid misunderstanding.
     }
    \label{fig:Stoner hsp}
\end{figure}

\begin{figure*}[htbp]
    \centering
    \begin{tabular}{lll}
    { (a) $\Sigma,\Delta(k;\bm{p})\rightarrow\Sigma,\Delta(k;\bm{0})$ } &{ (b) $\Sigma(k;\bm{p})\rightarrow\Sigma(k;\bm{0})$} &{ (c) $\Delta(k;\bm{p})\rightarrow\Delta(k;\bm{0})$}\\ 
     \includegraphics[width=0.3\textwidth,height=0.15\textheight]{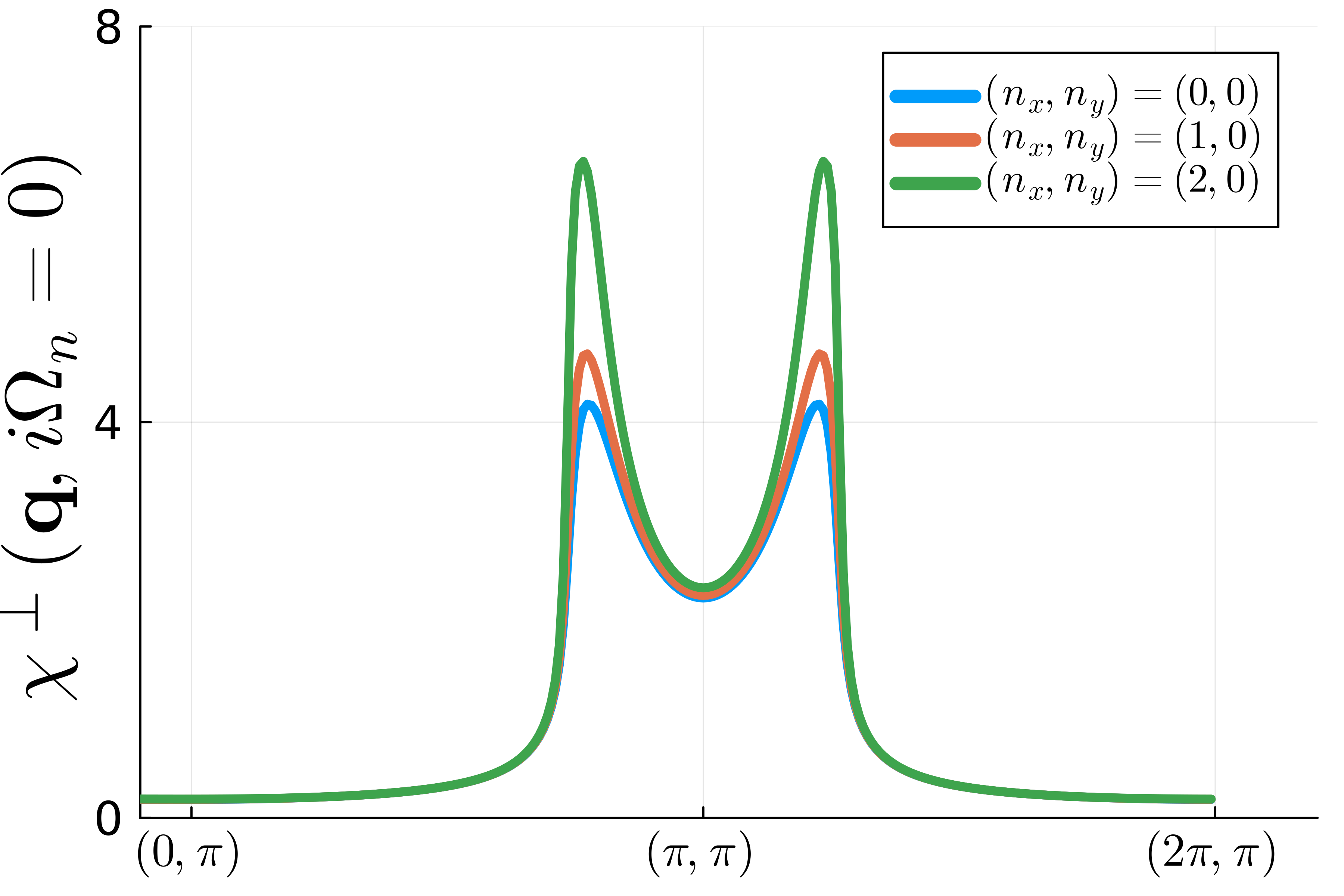} 
    &\includegraphics[width=0.3\textwidth,height=0.15\textheight]{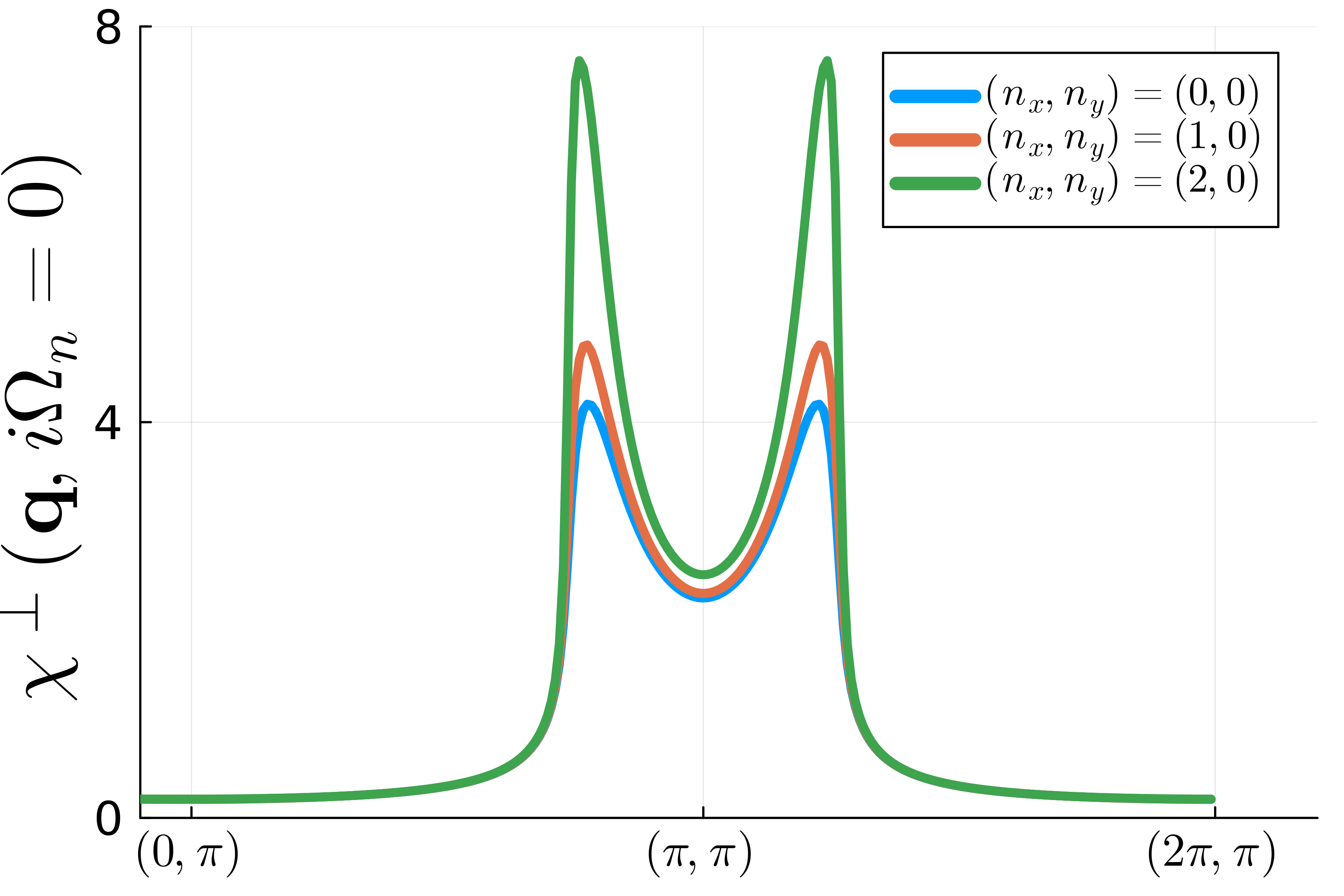} 
    &\includegraphics[width=0.3\textwidth,height=0.15\textheight]{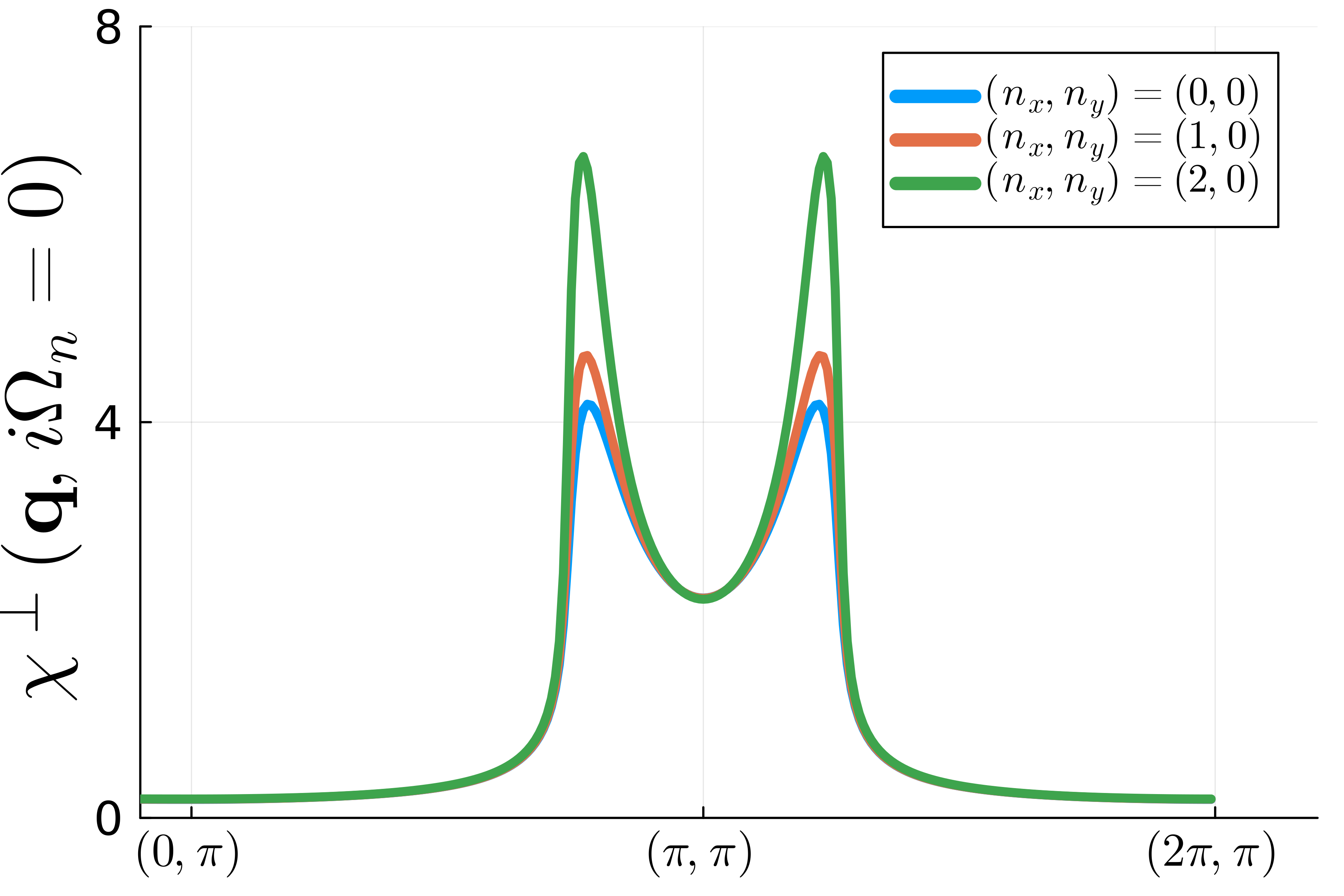} \\\\
    { (d) $\Sigma,\Delta(k;\bm{p})\rightarrow\Sigma,\Delta(k;\bm{0})$ } &{ (e) $\Sigma(k;\bm{p})\rightarrow\Sigma(k;\bm{0})$} &{ (f) $\Delta(k;\bm{p})\rightarrow\Delta(k;\bm{0})$}\\ 
     \includegraphics[width=0.3\textwidth,height=0.15\textheight]{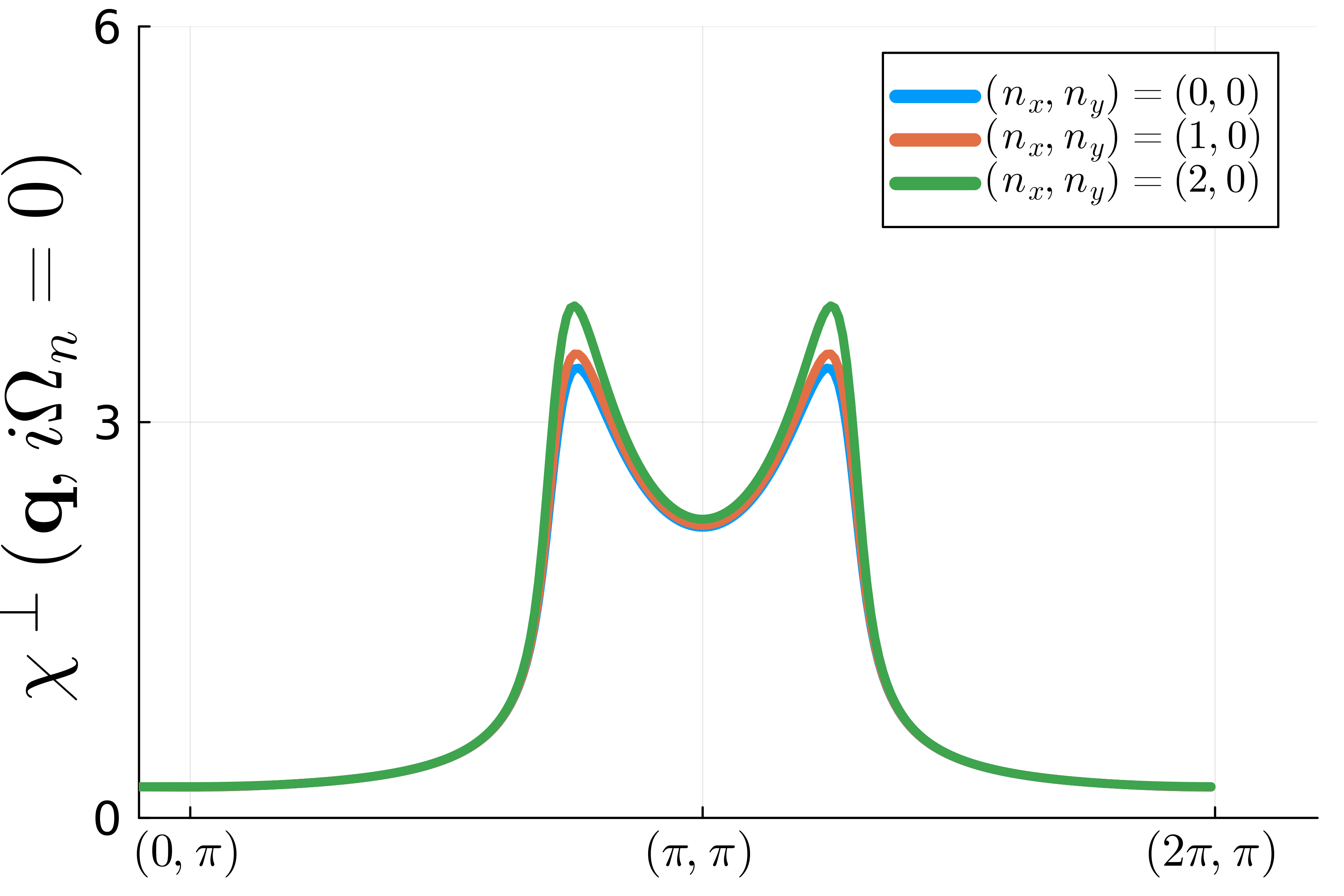} 
    &\includegraphics[width=0.3\textwidth,height=0.15\textheight]{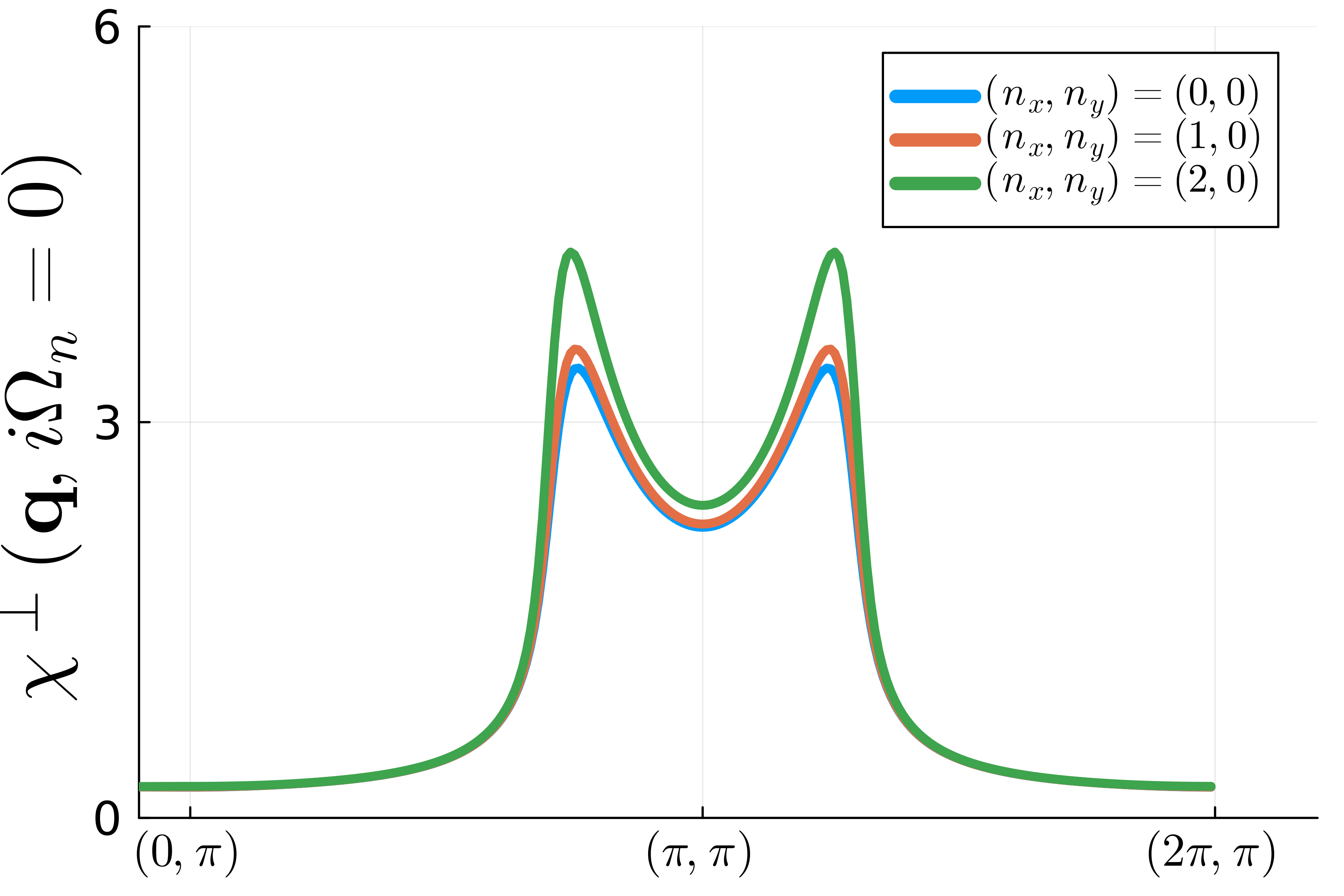}  
    &\includegraphics[width=0.3\textwidth,height=0.15\textheight]{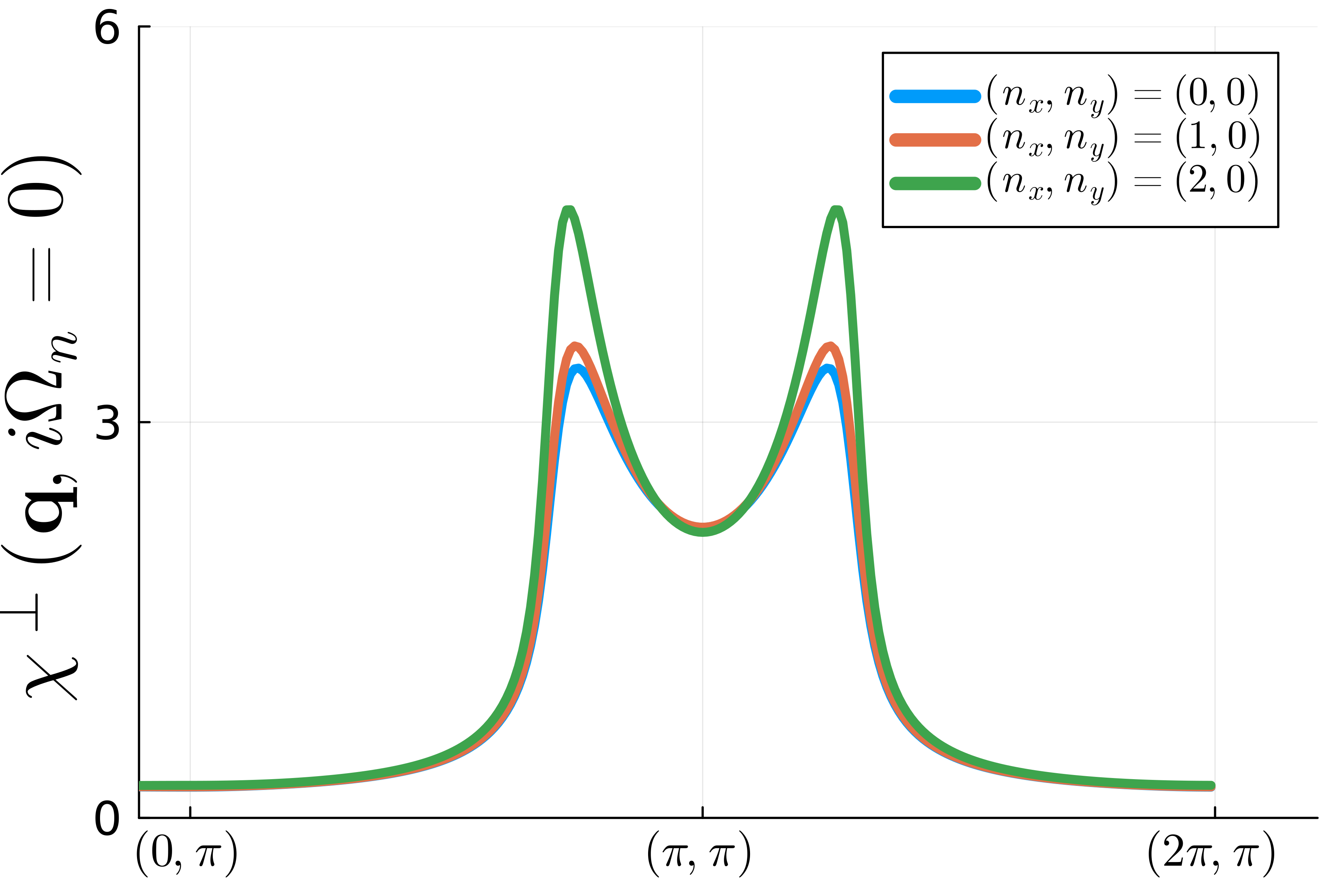}
    \end{tabular}
    \caption{Static spin susceptibility along the high symmetry momentum for (a)-(c) $(t',U)=(0.25,3.5)$ and (d)-(f) $(t',U)=(0.35,5.5)$.
    The green and orange lines show the spin susceptibility in the presence of the supercurrent along the [100]-axis, which are compared with the blue lines showing the results in the absence of the supercurrent. 
    (a), (d) $\bm{p}$-dependence of both normal and anomalous self-energies, $\Sigma(k;\bm{p})$ and $\Delta(k;\bm{p})$, is neglected. (b), (e) ${\bm p}$-dependence of the normal self-energy is neglected, while (c), (f) ${\bm p}$-dependence of the anomalous self-energy is neglected. 
%    The blue, orange, and green lines respectively correspond to the spin susceptibility for $(n_x,n_y)=(0,0),(1,0),$ and $(2,0)$.
%    The momentum of Cooper pairs, $(n_x,n_y)$, is used instead of the supercurrent.
     }
    \label{fig:spin sus fix}
\end{figure*}

Here, we show the supercurrent-induced AF order (SIAFO). 
In the absence of the supercurrent, the spin susceptibility shows four peaks near $\bm{q}\sim(\pi,\pi)$ [Fig.~\ref{fig:spin sus heatmap}(a)], indicating that the system is close to the AF quantum critical point. 
The maximum of spin susceptibility is enhanced by the supercurrent, as shown in Figs.~\ref{fig:spin sus heatmap}(b)-(e).
In particular, the spin susceptibility for $(n_x,n_y)=(2,0)$ shows a bimodal structure [Fig.~\ref{fig:spin sus heatmap}(e)] due to the symmetry breaking caused by the supercurrent along the [100]-axis. 
Consistent with the enhancement of the AF spin susceptibility, the Stoner factor is also enhanced by the supercurrent and overcomes the criterion $\alpha_S(\bm{q},i\Omega_n=0) = 0.98$ [Fig.~\ref{fig:Stoner hsp}]. This means that the supercurrent can induce the AF order. 
We have verified that SIAFO occurs for various parameter sets, such as $(t',U)=(0.3,5.0)$ and $(0.35,5.5)$. 
Therefore, it seems that SIAFO is a universal phenomenon in the Hubbard model.

We have confirmed that the spin susceptibility with the supercurrent along the $[100]$- and $[110]$-axes shows qualitatively the same behaviors, except for the bimodal structure for ${\bm p} \parallel (n_x,n_y) \parallel [100]$. Then, we focus on the supercurrent along the $[100]$-axis, which gives a quantitatively strong impact leading to the SIAFO. 
In the following, we discuss the mechanism of the SIAFO. 
As mentioned above, the SIAFO is a universal phenomenon in this model. However, the microscopic origin of the SIAFO depends on the parameters.
%We find that the microscopic origin depends on the parameters, although  occurs even for $(t',U)=(0.3,5.0)$ and $(0.35,5.5)$.
%Then, it seems that the SIAFO is a universal phenomenon in this model.

To clarify the main factors in the SIAFO, it is helpful to identify functions through which the supercurrent enhances antiferromagnetism.
For this purpose, we calculate the spin susceptibility while fixing some functions parameterized by Cooper pair's momentum $\bm{p}$ to those at $\bm{p}=\bm{0}$ by neglecting the supercurrent dependence.
The supercurrent changes the superconducting state through the energy dispersion of particles and holes $\varepsilon(\pm\bm{k}+\bm{p})$, the normal self-energy $\Sigma(k;\bm{p})$, and the anomalous self-energy $\Delta(k;\bm{p})$.
%Here, the functions are normal and anomalous self-energies,  and .
We carry out the following three calculations: (A) fixing both the normal and anomalous self-energies, (B) fixing the normal self-energy, and (C) fixing the anomalous self-energy.
For example, in case (A), the effects of supercurrent on superconductivity and magnetism arise only from the energy dispersion of free particles.

First, we discuss case (A), neglecting the supercurrent dependence of the normal and anomalous self-energies.
Then, only the energy dispersion, $\varepsilon(\pm\bm{k}+\bm{p})$ in Eq.~\eqref{eq:energy dispersion}, depends on the supercurrent.
In this condition, the supercurrent, or the energy dispersion modified by the supercurrent, increases the AF spin fluctuations for $(t',U)=(0.25,3.5)$, as shown in Fig.~\ref{fig:spin sus fix}(a), although it has little effect for $(t',U)=(0.35,5.5)$ [see Fig.~\ref{fig:spin sus fix}(d)].
The contrasting behaviors imply that Fermi surface nesting, 
which is enhanced by decreasing the next nearest neighbor hopping $t'$, plays a key role in SIAFO.

\begin{figure}[htbp]
    \centering
    \begin{tabular}{ll}
    (a) $(n_x,n_y)=(1,0)$  &(b) $(n_x,n_y)=(2,0)$\\ 
     \includegraphics[width=0.2\textwidth,height=0.14\textheight]{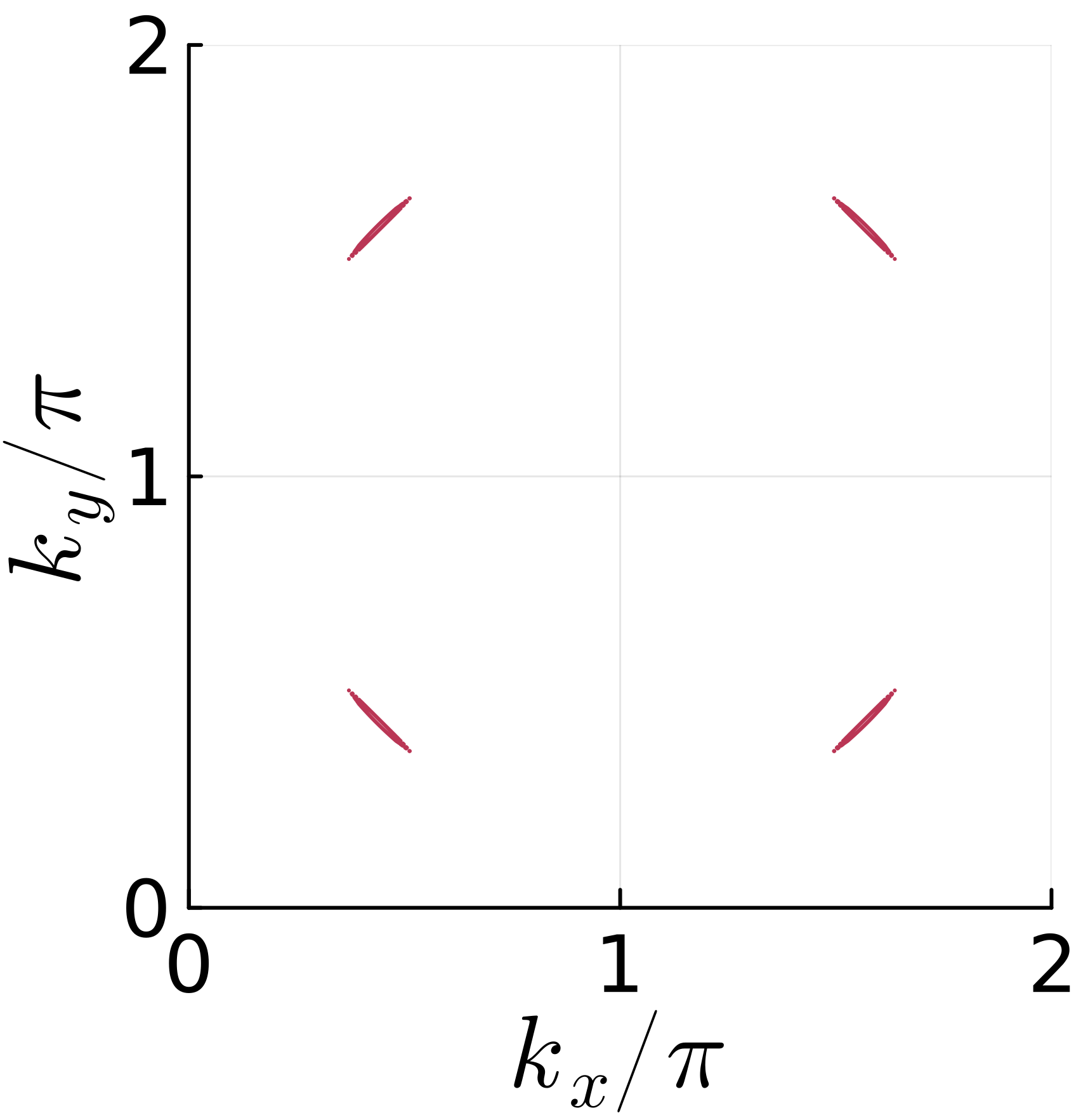} 
    &\includegraphics[width=0.2\textwidth,height=0.14\textheight]{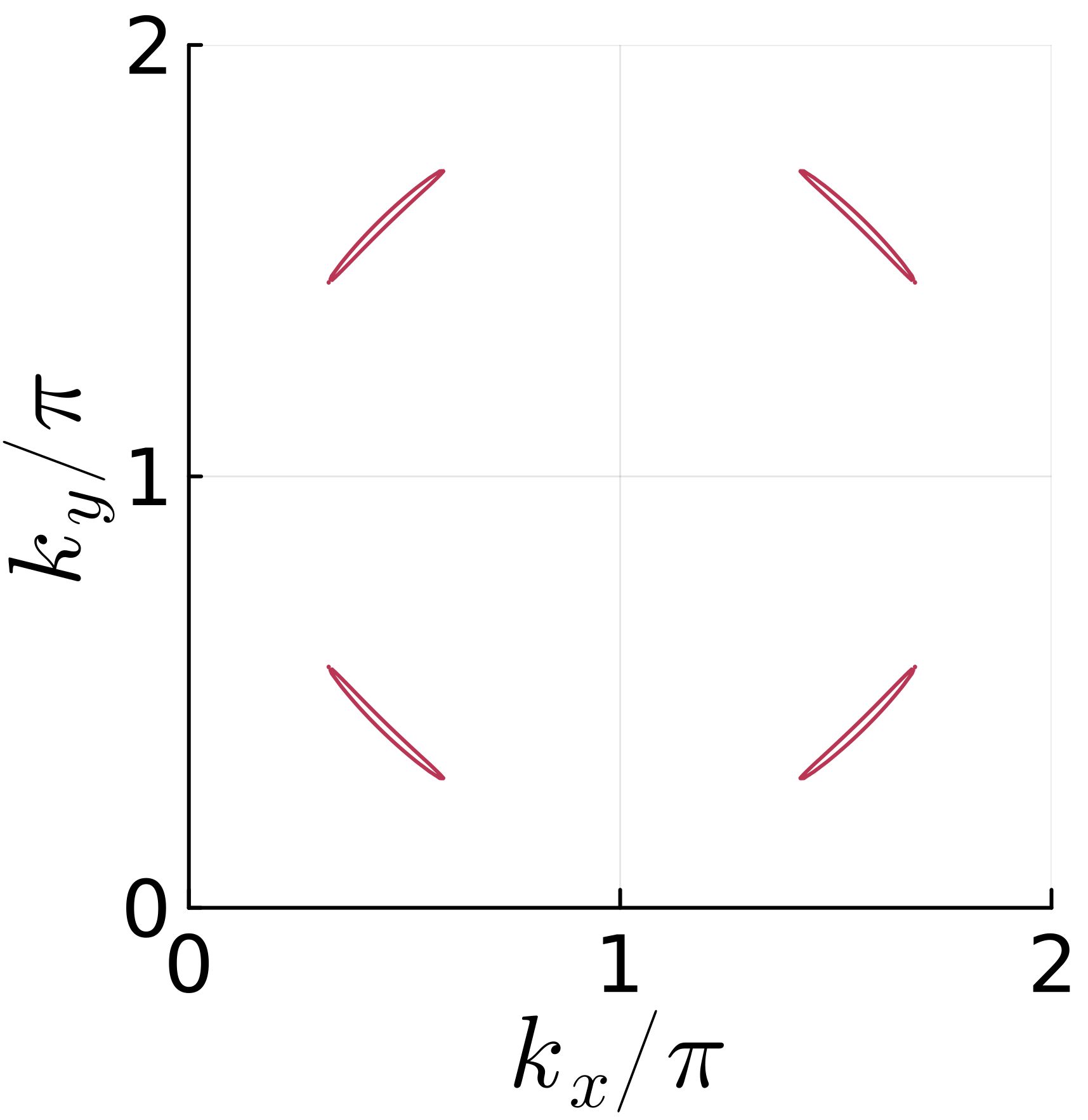} \\
    (c) $(n_x,n_y)=(1,0)$  &(d) $(n_x,n_y)=(2,0)$\\
     \includegraphics[width=0.2\textwidth,height=0.14\textheight]{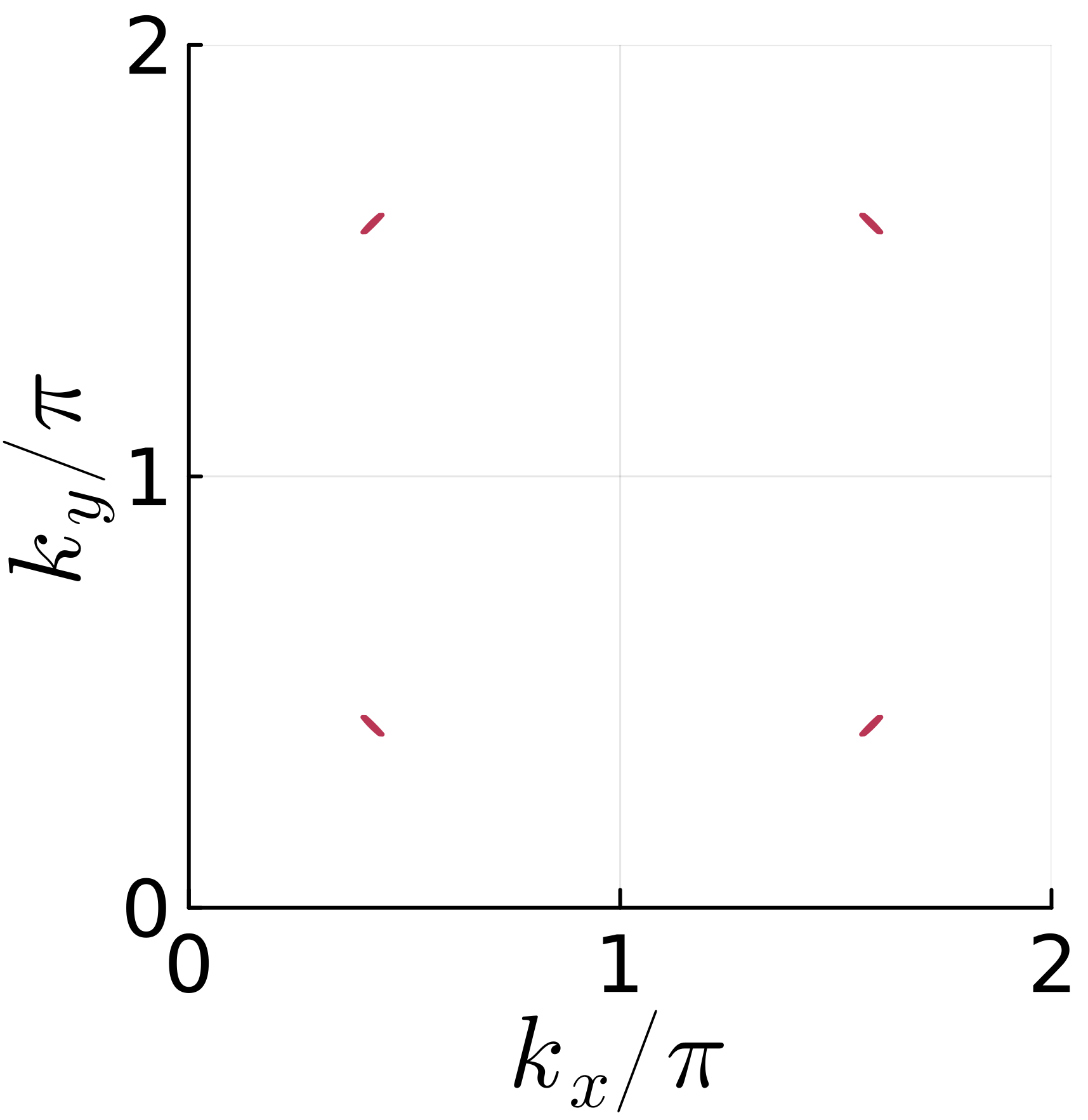} 
    &\includegraphics[width=0.2\textwidth,height=0.14\textheight]{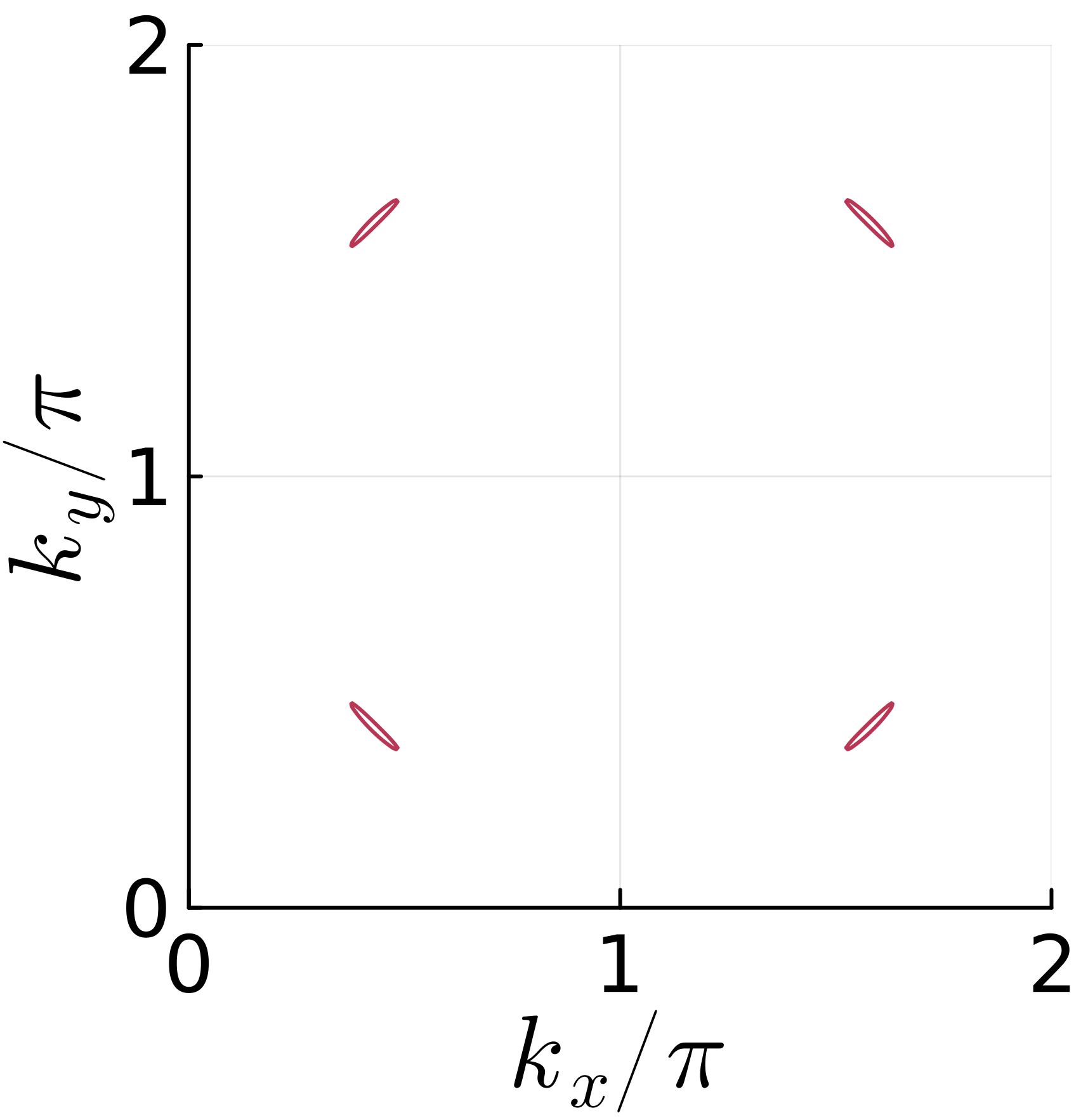}  
    \end{tabular}
    \caption{Supercurrent dependence of Bogoliubov Fermi surfaces for (a), (b) $(t',U)=(0.25,3.5)$ and (c), (d) $(t',U)=(0.35,5.5)$.
    The supercurrent is represented by the momentum of Cooper pairs, $(n_x,n_y)$.
     }
    \label{fig:BFS}
\end{figure}

The supercurrent modulates the energy dispersion and induces quasiparticle excitations around the nodal lines of $d$-wave superconductors~\cite{Franz-Millis}. This is the dominant effect in the calculation (A) and leads to a finite density of states at the Fermi level, called Bogoliubov Fermi surfaces (BFSs)~\cite{Agterberg2017,Brydon2018}.
%the BFSs can indicate the relationship between the supercurrent and the energy dispersion.
In the interacting systems, the BFSs are defined by the set of momentum satisfying the condition for the zero energy excitations, $\lambda_\pm(\bm{k})=0$, where $\lambda_\pm(\bm{k})$ is the eigenvalues of the following matrix,
\begin{align}
    \begin{pmatrix} \varepsilon(\bm{k}+\bm{p)}+\Sigma_\uparrow(\bm{k}) & \Delta(\bm{k})\\
     \Delta^\dagger(\bm{k}) & -\varepsilon(-\bm{k}+\bm{p})-\Sigma_\downarrow(\bm{k})  \end{pmatrix} .
\end{align}
Here, the static limit of the self-energy is evaluated by the following approximation for $A=\Sigma_\sigma$ and $\Delta$, 
\begin{align}
    A(\bm{k})=A(\bm{k},i\omega_n=0)\simeq\frac{A(\bm{k},i\pi T)+A(\bm{k},-i\pi T)}{2}, \label{eq:static function}
\end{align}
which is justified at low temperatures.
We show the BFSs induced by the supercurrent in Fig.~\ref{fig:BFS}.
The large BFSs are obtained for $(t',U)=(0.25,3.5)$ [Figs.~\ref{fig:BFS}(a) and \ref{fig:BFS}(b)], while the BFSs are small for $(t',U)=(0.35,5.5)$ [Figs.~\ref{fig:BFS}(c) and ~\ref{fig:BFS}(d)].
The size of the BFSs is well correlated with the increase in AF spin fluctuations in the calculation (A) [Figs.~\ref{fig:spin sus fix}(a) and \ref{fig:spin sus fix}(d)]. It is reasonable to consider that larger BFSs have a more significant impact on antiferromagnetism. 
Therefore, the supercurrent-enhanced AF spin fluctuation through modulation of energy dispersion is attributed to the emergence of BFSs, whose size depends on $t'$ and the Fermi surface nesting.
Note that %the precise relationship between the BFSs and magnetism remains unclear, although 
BFSs can appear not only by supercurrent but also through various origins, and spin fluctuations arising from the BFSs have been observed in $\mathrm{S}$-substituted $\mathrm{FeSe}$ by the nuclear magnetic resonance measurement~\cite{Yu2023}.

Second, we consider case (B), neglecting the supercurrent dependence of the normal self-energy.
Then, the supercurrent modifies the anomalous self-energy and the energy dispersion.
The supercurrent suppresses the anomalous self-energy and enhances AF spin fluctuations, as shown in Figs.~\ref{fig:spin sus fix}(b) and \ref{fig:spin sus fix}(e) compared to Figs.~\ref{fig:spin sus fix}(a) and \ref{fig:spin sus fix}(d).
This enhancement is qualitatively independent of $t'$ and is expected to be ubiquitous.

The supercurrent-enhanced AF spin fluctuations through the anomalous self-energy are naturally understood in the following way. 
The superconducting gap corresponds to the static anomalous self-energy and is reduced by the supercurrent because of the increase in the kinetic energy.
The superconducting gap suppresses quasiparticle excitations, and thus the imaginary part of the spin susceptibility at low energies.
Since the supercurrent reduces the superconducting gap, it weakens this suppression. Then, the Kramers-Kronig relation implies an increase in the real part of the static spin susceptibility. 
In fact, we have confirmed that the supercurrent slightly reduces the static anomalous self-energy, $\Delta_{\mathrm{max}}(\bm{k};\bm{p}\neq\bm{0})/\Delta_{\mathrm{max}}(\bm{k};\bm{p}=\bm{0})\sim0.98$ for the parameters in Fig.~\ref{fig:spin sus fix}.
Here, $\Delta_{\mathrm{max}}(\bm{k};\bm{p})$ is the maximum absolute value of the static anomalous self-energy.
Since the Stoner factor %for $(n_x,n_y)=(2,0)$ is almost 
is close to $1$ near the AF quantum critical point, even a small reduction in the superconducting gap significantly enhances the spin susceptibility and can cause the SIAFO. 
The mechanism discussed in this paragraph can be qualitatively understood as a result of competition between the superconducting and AF orders.

%Although the supercurrent also changes the $k$-dependence of the anomalous self-energy, $\Delta(k)$, we have verified that this alteration does not affect the AF fluctuations as much as the alternation of the amplitude of the anomalous self-energy.

Third, we discuss case (C), neglecting the supercurrent dependence of the anomalous self-energy.
Then, the normal self-energy and the energy dispersion are affected by the supercurrent.
AF spin fluctuations are almost unaffected by the supercurrent dependence of the normal self-energy for $(t',U)=(0.25,3.5)$ [compare Fig.~\ref{fig:spin sus fix}(c) with \ref{fig:spin sus fix}(a)], while manifestly enhanced for $(t',U)=(0.35,5.5)$ [compare Fig.~\ref{fig:spin sus fix}(f) with \ref{fig:spin sus fix}(d)]. 
The normal self-energy represents the correlation effects beyond the mean-field theory, and we see that it affects the SIAFO in a different way between the two cases with strong or weak Fermi surface nesting. 
The correlation effects (do not) enhance the SIAFO when the Fermi surface nesting is relatively weak (strong). This is in contrast to the role of BFSs. Thus, although SIAFO is an ubiquitous phenomenon in $d$-wave superconductors near the AF quantum critical point, the microscopic mechanism depends on the band structure.

In addition, we refer to the spin susceptibility obtained without fixing normal and anomalous self-energies, that is, the spin susceptibility obtained by the FLEX approximation.
Some of the results are shown in Fig.~\ref{fig:spin sus heatmap}. In this case, the effects involved in cases (A), (B), and (C) cooperatively contribute to SIAFO.
For $(t',U)=(0.25,3.5)$, the dominant contribution to SIAFO originates from BFSs and anomalous self-energy.
Then, the FLEX approximation results in almost the same spin susceptibility as in Fig.~\ref{fig:spin sus fix}(b).
On the other hand, for $(t',U)=(0.35,5.5)$, normal and anomalous self-energies give dominant contributions to SIAFO.
Hence, the maximum value of the spin susceptibility exceeds those in Figs.~\ref{fig:spin sus fix}(e) and \ref{fig:spin sus fix}(f) and reaches about $5.3$.

Parenthetically, we comment on a previous theoretical study of SIAFO~\cite{Takashima2018}.
The formulation in Ref.~\onlinecite{Takashima2018} is different from ours in the following three points:
$(1)$ a mean-field theory for magnetism is adopted with a superconducting gap being phenomenologically fixed, 
$(2)$ $s$-wave superconductivity is assumed, 
and $(3)$ the next nearest hopping is set to zero, $t'=0$.
%where the mean field is not a superconducting gap but a spin density, and thus the superconducting gap is just a parameter.
%The results of the study also have shown the SIAFO which has almost the same structure and supercurrent dependence as our results in Fig.~\ref{fig:spin sus heatmap}.
Due to the point $(1)$, the contributions of self-energies evaluated in our calculations (B) and (C) are neglected. Therefore, the SIAFO studied in Ref.~\onlinecite{Takashima2018}  seems to arise from the BFSs as in our calculation (A).
%Although the point (2) presents that the superconductivity has no nodes, the supercurrent can induce the BFSs because the superconducting gap being just a parameter enables the supercurrent to be much larger compared to the superconducting gap.
Because the $s$-wave superconducting state has no gap node, BFSs appear only when the supercurrent is sufficiently large to suppress the superconducting gap. The stability of the superconducting state is not guaranteed under such a large supercurrent~\cite{Daido2022}. 
In contrast, in our calculation, a small supercurrent can induce BFSs in gapless $d$-wave superconductors, which favor SIAFO. The stability of the superconducting state is demonstrated by our self-consistent calculations. Corrections to the self-energies are also cooperative. 
These comparisons of Ref.~\onlinecite{Takashima2018} and our study highlight the advantages of strongly correlated $d$-wave superconductors for SIAFO.
%Thus, the large supercurrent can induce the AF order even within $S$-wave superconductivity and the mean field theory.
%However, a larger $t'$ than zero (see point $(1)$) shrinks the size of the BFSs and prevents the mean field theory from describing the SIAFO.

Finally, we summarize the three calculations (A), (B), and (C).
The supercurrent expands the BFSs, reduces the superconducting gap, and changes the strong correlation effects.  %These effects appear through the energy dispersion, the anomalous self-energy, and the normal self-energy. 
These effects of the supercurrent enhance the AF spin fluctuations and trigger the AF order. The quantitative importance of each contribution depends on the next nearest neighbor hopping $t'$ that controls the Fermi surface nesting.
When the Fermi surface nesting is relatively strong ($t'=0.25$), the dominant contributions come from the BFSs and the superconducting gap.
On the other hand, when the Fermi surface nesting is relatively weak ($t'=0.35$), the contributions of the correlation effects and the superconducting gap are dominant.
%We have verified that when the Fermi surface nesting is medium ($t'=0.3$), the BFSs, the superconducting gap, and the strong correlation effects all contribute to the SIAFO.
We confirmed that the relative importance of these contributions to SIAFO gradually changes as the Fermi surface is altered. 
We emphasize that the correlation effects, such as the lifetime and mass renormalization of quasiparticles, may play an essential role for SIAFO, although they go beyond the mean-field theory and the quasiclassical theory.
In other words, the SIAFO offers a promising means of dynamically controlling superconductivity and magnetism in SCES via supercurrent. 
%arising from the strong correlation effects 
%would be an important bridge between the supercurrent and superconductivity in SCES. 

\subsection{Spin-triplet Cooper pair generation}
\label{sec:superconductivity}

\begin{figure}[htbp]
    \centering
    \begin{tabular}{cc}
    (a) $(n_x,n_y)=(1,1)$&(b) $(n_x,n_y)=(2,2)$ \\ 
     \includegraphics[width=0.21\textwidth,height=0.14\textheight]{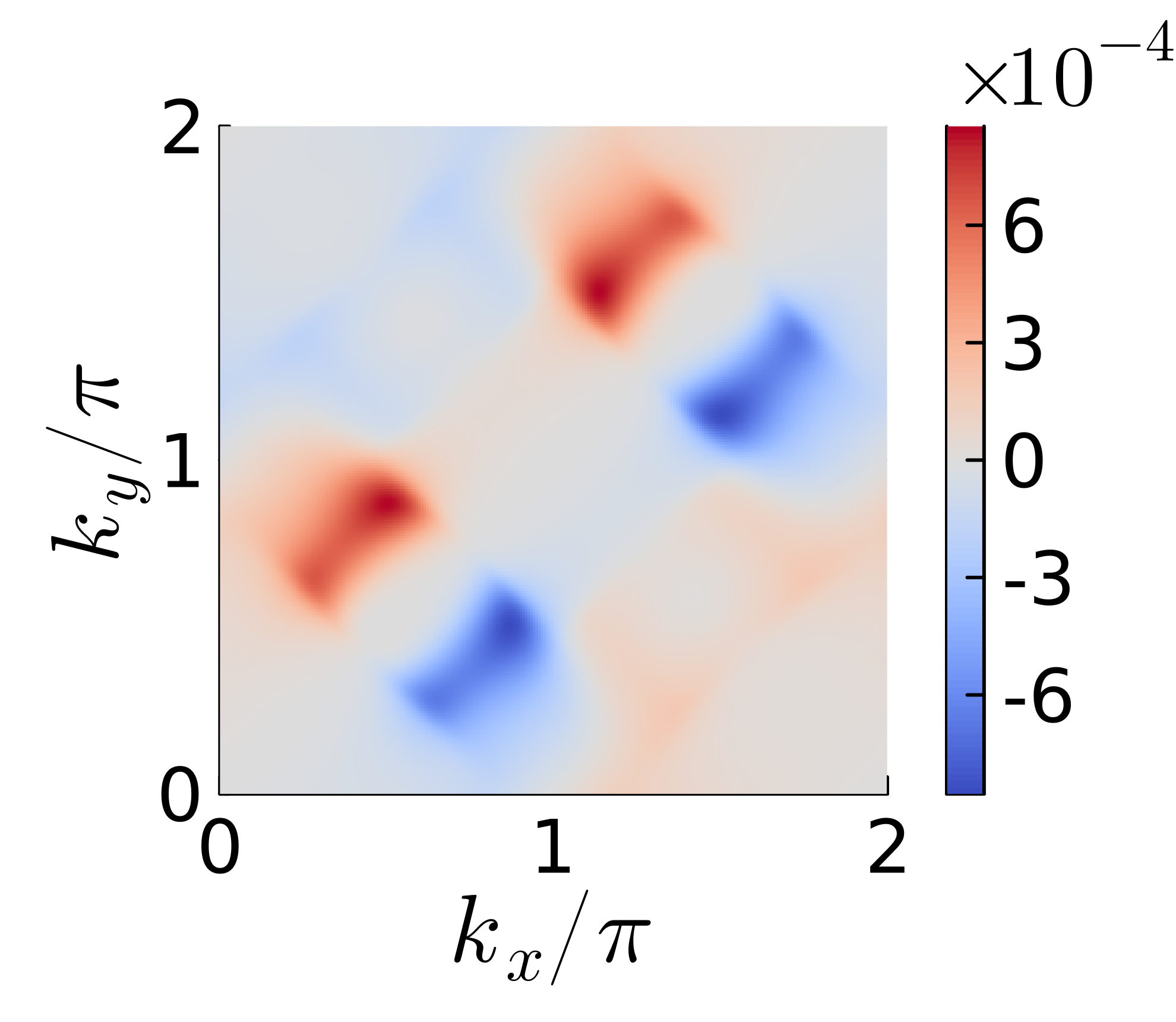} 
    &\includegraphics[width=0.21\textwidth,height=0.14\textheight]{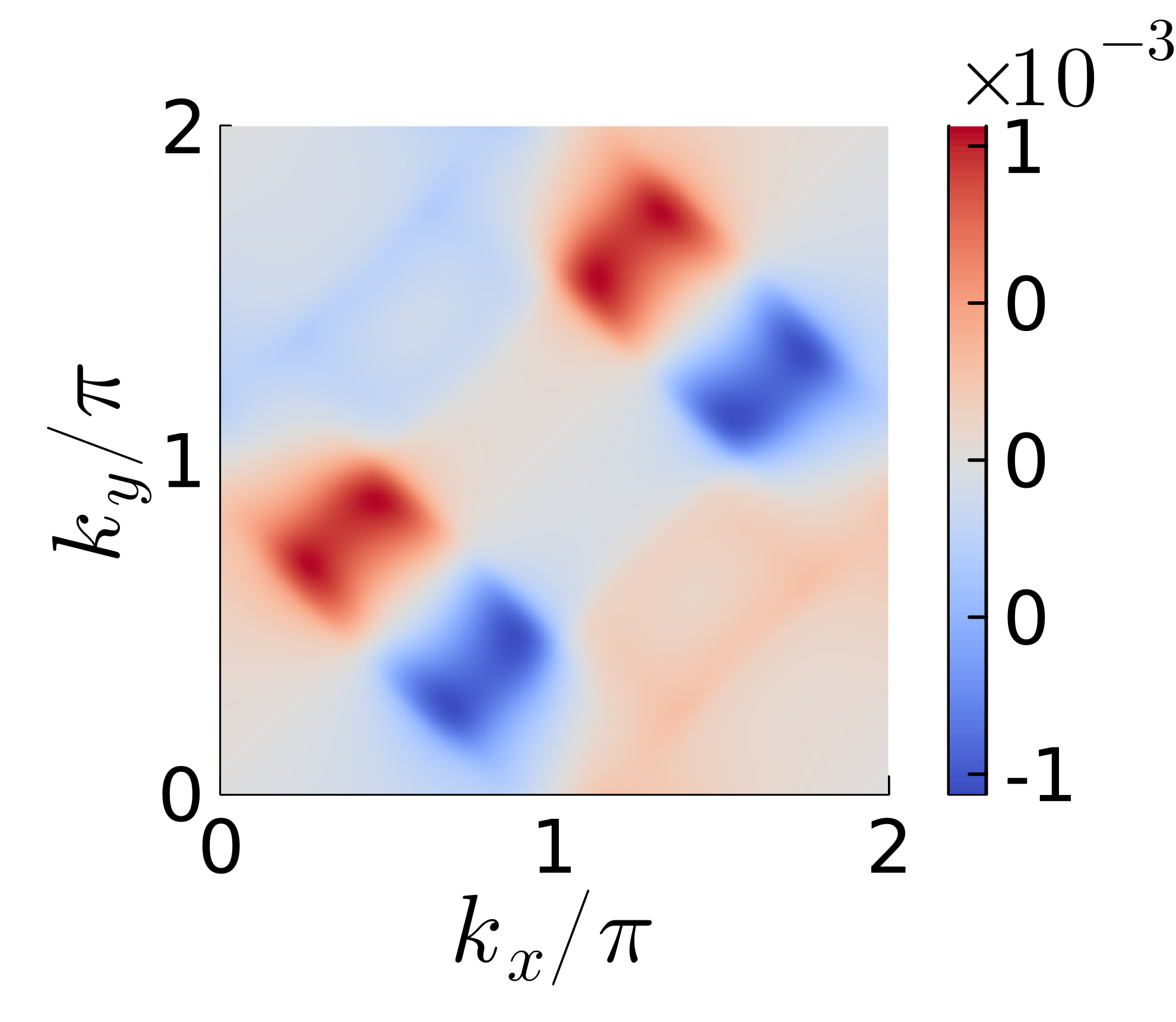} \\
    (c) $(n_x,n_y)=(1,0)$&(d) $(n_x,n_y)=(2,0)$\\
     \includegraphics[width=0.21\textwidth,height=0.14\textheight]{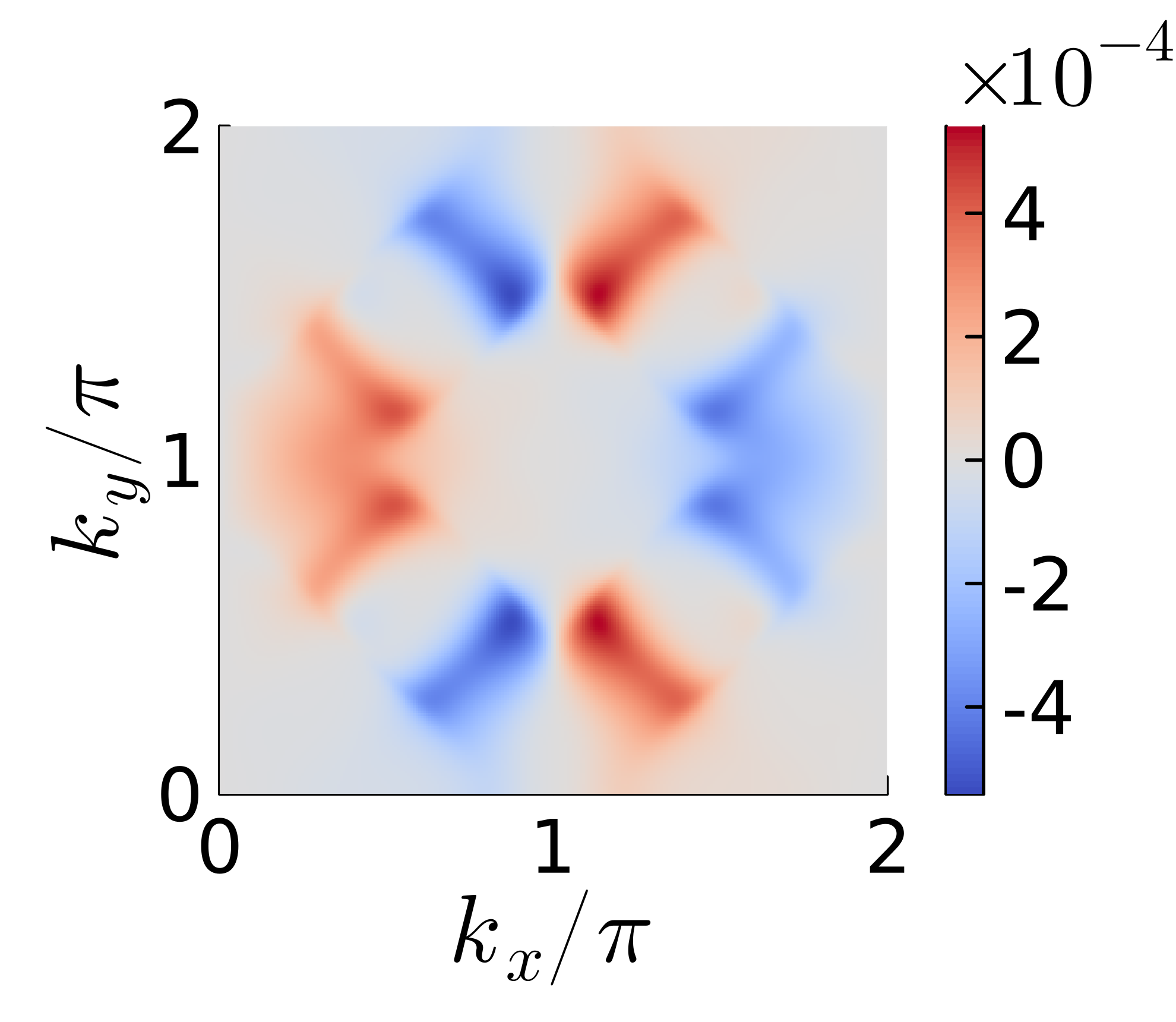}
    &\includegraphics[width=0.21\textwidth,height=0.14\textheight]{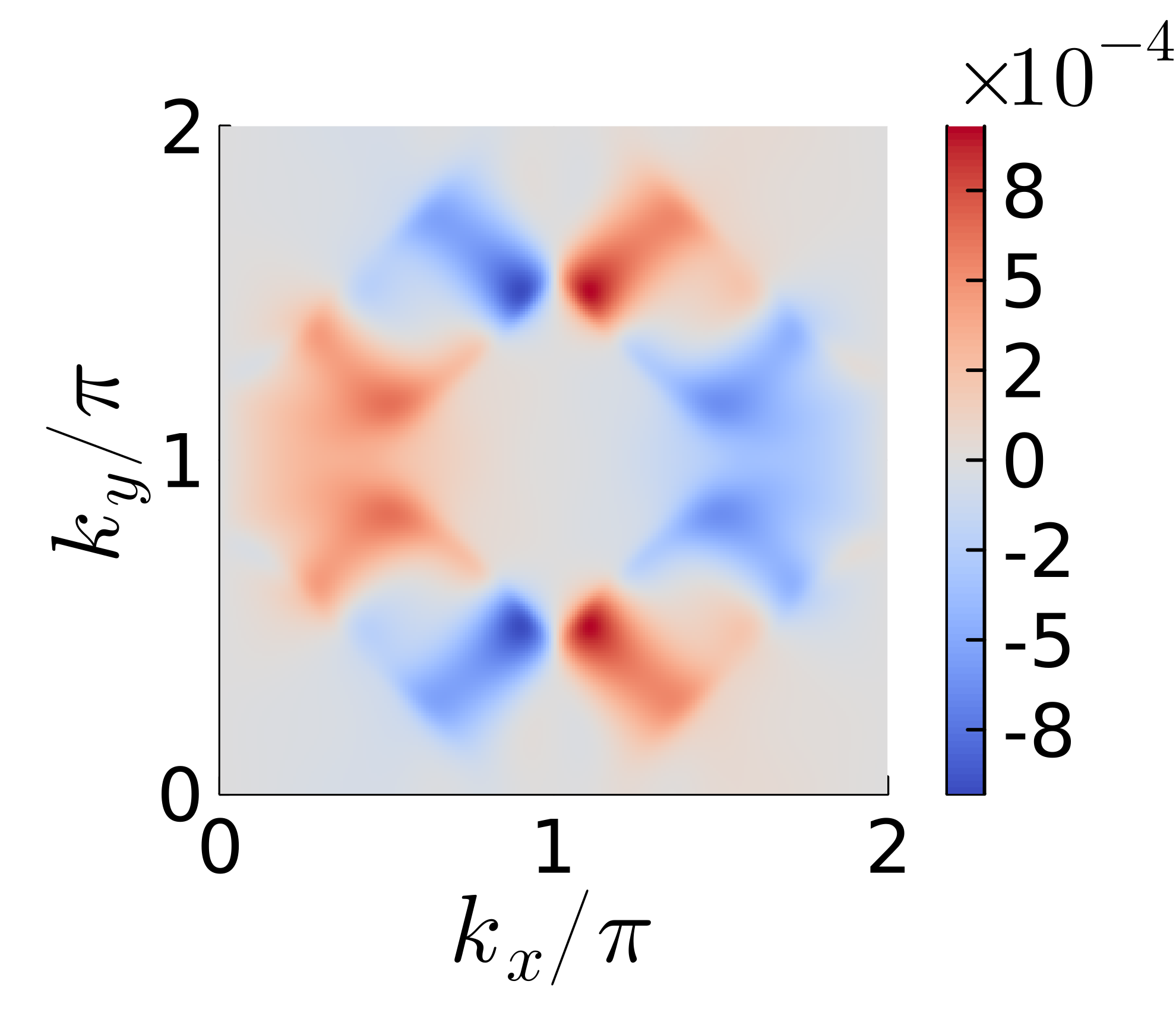} 
    \end{tabular}
    \caption{Spin-triplet component, $\Psi_t(\vb{k})$, of the static anomalous self-energy that corresponds to the pair potential. We set the parameters $(t',U)=(0.35,5.5)$ and assume the magnetic field $h=0.005$.
    The supercurrent is represented by the momentum of Cooper pairs, $(n_x,n_y)$.
    Note that the spin-triplet component vanishes in the absence of the supercurrent, $(n_x,n_y) = (0,0)$.}
    \label{fig:triplet}
\end{figure}

In this subsection, we demonstrate spin-triplet Cooper pairs induced by the supercurrent.
The supercurrent breaks the space inversion symmetry and allows for parity mixing of superconductivity.
However, at zero magnetic field $h=0$, %drops the distinction between up and down spin derived from the Zeeman term in Eq.~\ref{eq:Hubbard model}, 
the isotropy in spin space prohibits mixing of spin-singlet even-parity and spin-triplet odd-parity Cooper pairs, and thus the spin-triplet pairing does not occur.
Then, cooperation of supercurrent and magnetic field generates spin-triplet Cooper pairs.

Spin-singlet and spin-triplet components, $\Psi_s$ and $\Psi_t$, are defined by the anomalous self-energy as
\begin{align}
    \Psi_s(k)&=\frac{\Delta_{\uparrow\downarrow}(k)-\Delta_{\downarrow\uparrow}(k)}{2}, \label{eq:singlet} \\
    \Psi_t(k)&=\frac{\Delta_{\uparrow\downarrow}(k)+\Delta_{\downarrow\uparrow}(k)}{2} .
    \label{eq:triplet}
\end{align}
The static anomalous self-energy is decomposed into the spin-singlet and spin-triplet components by using Eqs.~\eqref{eq:singlet} and \eqref{eq:triplet} with Eq.~\eqref{eq:static function}.
Figure~\ref{fig:triplet} shows that the spin-triplet component emerges with a finite supercurrent and the gap structure depends on the direction of the supercurrent.
We have also confirmed that the spin-triplet component becomes larger as the magnetic field becomes larger.
%in Fig.~\ref{fig:triplet}. 
%which means that the triplet component also increases with field.
%In addition, we have verified that the supercurrent and field increase the triplet component even for $t'=0.25$, $0.35$.
Note that the spin-singlet $d$-wave component is always dominant, that is, $|\Psi_s(\vb{k})|\sim\mathcal{O}(10^{-1})\gg|\Psi_t(\vb{k})|$, as is naturally expected because the pairing interaction is mediated by AF spin fluctuations. 
However, our results reveal that the supercurrent can generate spin-triplet Cooper pairs with $d$-vector parallel to the magnetic field via an effective interaction mediated by AF spin fluctuations. 
Therefore, a supercurrent is accompanied by a spin-triplet Cooper pair current in SCES.

\section{conclusion}
\label{sec:conclusion}

In this paper, we studied the superconducting state with a supercurrent in the two-dimensional Hubbard model.
It has been shown that an antiferromagnetic order and spin-triplet Cooper pairs can be induced by a supercurrent. %have been revealed in the two-dimensional Hubbard model by using a fluctuation exchange approximation with a finite momentum of Cooper pairs.
Our calculations based on the FLEX approximation reveal electron correlation effects beyond the mean-field theory;
the supercurrent-induced antiferromagnetic order results from appearance of Bogoliubov Fermi surfaces, suppression of the superconducting gap, and modification of correlation effects. 
The results in this paper demonstrate that the supercurrent can be a control parameter of superconductivity and magnetism in SCES.
%In particular, the SIAFO derived from the strong correlation effects is the phenomenon beyond the mean field theory and the quasiclassical theory, which suggests that the supercurrent is an important parameter in strongly correlated electron systems (SCES).
The prediction can be verified in various systems, including high-$T_{\rm c}$ cuprate superconductors, organic superconductors, and heavy fermion superconductors.
Our findings would extend the research of dynamical superconductivity and superconducting spintronics to SCES, a fertile ground for quantum phenomena.
%Moreover, the possibility of the contribution to opening a new way of manipulating quantum phases is an exciting proposition.

\begin{acknowledgments}
The authors thank for fruitful discussions with Taisei Kitamura. 
This work was supported by JSPS KAKENHI (Grant Numbers JP22H01181, JP22H04933,
JP23K17353, JP23K22452, JP24K21530, JP24H00007, 24KJ1475, JP25H01249).
\end{acknowledgments}

\bibliography{ref}

\end{document}